\documentstyle[epsfig,11pt,aaspp4]{article}

\newcommand\uzeroone{IR00509$+$1225\,}
\newcommand\uzerotwo{IR02364$-$4751\,}
\newcommand\uzerothree{IR06035$-$7102\,}
\newcommand\uzerofour{IR06206$-$6315\,}
\newcommand\uzerofive{IR09320$+$6134\,}
\newcommand\uzerosix{IR11095$-$0238\,}
\newcommand\uzeroseven{IR12112$+$0305\,}
\newcommand\uzeroeight{IR13428$+$5608\,}
\newcommand\uzeronine{IR14348$-$1447\,}
\newcommand\uonezero{IR14378$-$3651\,}
\newcommand\uonefour{IR19254$-$7245\,}
\newcommand\uonefive{IR19297$-$0406\,}
\newcommand\uonesix{IR20087$-$0308\,}
\newcommand\uoneseven{IR20100$-$4156\,}
\newcommand\uoneeight{IR20414$-$1651\,}
\newcommand\uonenine{IR20551$-$4250\,}
\newcommand\utwozero{IR21130$-$4446\,}
\newcommand\utwoone{IR22491$-$1808\,}
\newcommand\utwotwo{IR23128$-$5919\,}
\newcommand\utwothree{IR23230$-$6926\,}
\newcommand\utwofour{IR13144$+$2356\,}
\newcommand\utwofive{IR13342$+$3932\,}
\newcommand\utwosix{IR13442$+$2321\,}
\newcommand\utwoseven{IR13539$+$2920\,}
\newcommand\utwoeight{IR14060$+$2919\,}
\newcommand\utwonine{IR14170$+$4545\,}
\newcommand\uthreezero{IR14202$+$2615\,}
\newcommand\uthreeone{IR14312$+$2825\,}
\newcommand\uthreetwo{IR14575$+$3256\,}
\newcommand\uthreethree{IR00150$+$4937\,}
\newcommand\uthreefour{IR12490$-$1009\,}
\newcommand\uthreesix{IR13469$+$5833\,}
\newcommand\uthreeeight{IR15413$-$0959\,}
\newcommand\uthreenine{IR16455$+$4553\,}
\newcommand\ufourone{IR16541$+$5301\,}
\newcommand\ufourtwo{IR17179$+$5444\,}
\newcommand\ufourfour{IR18580$+$6527\,}
\newcommand\ufournine{IR09425$+$1751\,}
\newcommand\ufivezero{IR09427$+$1929\,}
\newcommand\ufivefour{IR16007$+$3743\,}
\newcommand\ufivesix{IR23365$+$3604\,}
\newcommand\ufiveseven{IR00204$+$1029\,}
\newcommand\ufivenine{IR00275$-$2859\,}
\newcommand\usixzero{IR00461$-$0728\,}
\newcommand\usixfive{IR02054$+$0835\,}
\newcommand\usixeight{IR03538$-$6432\,}
\newcommand\usixnine{IR04384$-$4848\,}
\newcommand\usevenzero{IR04413$+$2608\,}
\newcommand\usevenone{IR05120$-$4811\,}
\newcommand\useventwo{IR05233$-$2334\,}
\newcommand\useventhree{IR06268$+$3509\,}
\newcommand\usevenfour{IR06361$-$6217\,}
\newcommand\usevenfive{IR06561$+$1902\,}
\newcommand\usevensix{IR07381$+$3215\,}
\newcommand\usevenseven{IR08201$+$2801\,}
\newcommand\useveneight{IR10026$+$4347\,}
\newcommand\usevennine{IR10558$+$3845\,}
\newcommand\ueightzero{IR10579$+$0438\,}
\newcommand\ueightone{IR12108$+$3157\,}
\newcommand\ueighttwo{IR12202$+$1646\,}
\newcommand\ueightthree{IR13352$+$6402\,}
\newcommand\ueightfive{IR14337$-$4134\,}
\newcommand\ueightsix{IR16159$-$0402\,}
\newcommand\ueightseven{IR17431$-$5157\,}
\newcommand\ueighteight{IR17463$+$5806\,}
\newcommand\uninezero{IR20037$-$1547\,}
\newcommand\unineone{IR20109$-$3003\,}
\newcommand\uninetwo{IR20176$-$4756\,}
\newcommand\uninethree{IR20253$-$3757\,}
\newcommand\uninefour{IR20314$-$1919\,}
\newcommand\uninefive{IR20507$-$5412\,}
\newcommand\uninesix{IR21547$-$5823\,}
\newcommand\unineseven{IR23140$+$0348\,}
\newcommand\unineeight{IR23220$+$2919\,}
\newcommand\ubzero{IR04024$-$8303\,}
\newcommand\ubone{IR05116$+$7745\,}
\newcommand\ubfour{IR06487$+$2208\,}
\newcommand\ubfive{IR07246$+$6125\,}
\newcommand\ubseven{IR08344$+$5105\,}
\newcommand\ubeight{IR08509$-$1504\,}
\newcommand\ubnine{IR09039$+$0503\,}
\newcommand\uctwo{IR10122$+$4943\,}
\newcommand\ucthree{IR11087$+$5351\,}
\newcommand\ucfive{IR19561$-$4756\,}
\newcommand\ucseven{IR00060$-$1543\,}
\newcommand\ucnine{IR00105$-$0139\,}
\newcommand\udzero{IR00161$-$0850\,}
\newcommand\udtwo{IR00335$-$2732\,}
\newcommand\udfour{IR00589$-$0352\,}
\newcommand\udfive{IR01031$-$2255\,}
\newcommand\ufzero{IR02459$-$0233\,}
\newcommand\ufthree{IR22206$-$2715\,}
\newcommand\uffive{IR22546$-$2637\,}
\newcommand\ufsix{IR23146$-$1116\,}
\newcommand\ufseven{IR23242$-$0357\,}
\newcommand\ufeight{IR23410$+$0228\,}
\newcommand\ufnine{IR23515$-$2421\,}

\begin{document}

\title{Statistical properties of ultraluminous IRAS galaxies\\
from an HST imaging survey}

\author{J. Cui}
\affil{Beijing Astronomical Observatory, Chinese Academy of Sciences\\
	Electronic mail: cuijun@bac.pku.edu.cn}
\author{X.-Y. Xia}
\affil{Dept. of Physics, Tianjin Normal University\\
	Electronic mail: xyxia@bac.pku.edu.cn}
\author{Z.-G. Deng}
\affil{Dept. of Physics, Graduate School, Chinese Academy of Sciences\\
	Electronic mail: dzg@bac.pku.edu.cn}
\author{S. Mao}
\affil{Jodrell Bank Observatory, University of Manchester\\
	Electronic mail: smao@jb.man.ac.uk}
\author{Z.-L. Zou}
\affil{Beijing Astronomical Observatory, Chinese Academy of Sciences\\
	Electronic mail: zzl@class1.bao.ac.cn}

\begin{abstract}

We perform photometric measurements on a large
{\it HST} snapshot imaging survey sample of 97
ultraluminous infrared galaxies (ULIRGs). We select
putative nuclei from bright clumps in all the sample targets, mainly
based on a quantitative criterion of {\it I}-band luminosity as
well as the global and local morphological information.
All the sources are then classified into three categories with
multiple, double and single nucleus/nuclei. The resultant
fractions of multiple, double and single nucleus/nuclei ULIRGs
are 18\%, 39\% and 43\%, respectively. This supports the
multiple merger scenario as a possible origin of ULIRGs,
in addition to the commonly-accepted pair merger model.
Further statistical studies indicate that the AGN fraction
increases from multiple (36\%) to double (65\%) and then to
single (80\%) nucleus/nuclei ULIRGs. For the single nucleus
category, there is a high luminosity tail in the luminosity
distribution, which corresponds to a Seyfert~1/QSO excess.
This indicates that active galactic nuclei tend to appear at
final merging stage. For multiple/double nuclei galaxies, we
also find a high fraction of very close nucleus pairs (e.g., 3/4
for those separated by less than 5\,kpc). This strengthens
the conclusion that systems at late merging phase
preferentially host ULIRGs.

\end{abstract}


\section{Introduction}

For more than ten years, much effort has been made 
to understand the properties of ultraluminous infrared
galaxies (hereafter ULIRGs, see Sanders \& Mirabel 1996
for a review), and it is now widely accepted that this
population of objects represents an important evolutionary
stage triggered by strong galaxy interactions/mergers,
especially major mergers between gas-rich spirals with 
mass ratio smaller than 3:1 (e.g., Bendo \& Barnes 2000).
These ULIRGs may represent an important stage in the
formation of QSOs and powerful radio galaxies, as well
as an essential step in the formation of elliptical
galaxies (e.g., Sanders {\it et al.} 1988a; Melnick
\& Mirabel 1990). Although the nature of ULIRGs is
believed to be understood in general terms, it is still
not clear whether these objects originate from the
interactions/mergers of galaxy pairs or galaxy groups.

On the theoretical side, numerical simulations have 
successfully reproduced global and even some fine
structural properties of elliptical-like galaxies as
the results of pair mergers (e.g., Toomre \& Toomre
1972; White 1978; Farouki \& Shapiro 1982; Negroponte
\& White 1983; Barnes 1992; Hernquist 1992; Heyl
{\it et al.} 1994; Barnes \& Hernquist 1996; Weil
\& Hernquist 1996; Walker {\it et al.} 1996; Mihos \&
Hernquist 1994; Springel \& White 1998). However, the
pair merger model can not account for all the observed
properties of elliptical galaxies, which leads people
to consider the multiple merger scenario (e.g., Mamon
1987; Barnes 1984, 1985, 1989; Schweizer 1989; Weil
\& Hernquist 1996). In fact, numerical simulations
of multiple mergers do produce remnants different from
those produced by pair mergers in both morphological
and kinematic properties (Weil \& Hernquist 1996).
On the other hand, there is increasing observational
evidence to support the multiple merger scenario. For
example, the co-existence of three OH maser components
in the typical ULIRG Arp~220 is of possible multiple
merger origin (Taniguchi \& Shioya 1998). And Lipari
{\it et al.} (1999) reported detailed evidence for
multiple merger in the luminous infrared galaxy NGC~3256
based on high resolution imaging by {\it HST} and
{\it ESO-NTT}. Furthermore, Borne {\it et al.} (1999a,
1999b) carried out a thorough ULIRGs imaging survey by
{\it HST} WFPC2/NICMOS, of which they displayed about
20 ULIRGs either with multiple nuclei or representing
interacting groups (Borne {\it et al.} 2000). Based
on their results, Borne {\it et al.} (2000) suggested
that a possible progenitor of the ULIRG population is
compact groups of galaxies.

In optical and near infrared images of luminous merging
galaxies, such as ULIRGs, there always exist many kpc
or sub-kpc scale intense star-forming regions (Lutz
1991; Ashman \& Zepf 1992; Holtzman {\it et al.} 1992;
Zepf \& Ashman 1993; Surace \& Sanders 1999). It has
often been argued that these star clusters may form
during violent collisions between gas clouds induced by
tidal interactions (Hutchings 1995), and that their
formation may be linked directly to intense starbursts
occurring at the very centers of ULIRGs (Taniguchi {\it
et al.} 1998). Compared with numerous star-forming
regions in nearby interacting systems or starburst
galaxies such as NGC~4038/9 (Whitmore \& Schweizer 1995),
NGC~3310 and NGC~2415 (Gallagher {\it et al.} 2000),
star clusters in ULIRGs are generally more luminous and
massive, with bolometric luminosity, $L_{\rm{bol}}$,
sometimes ranging up to $2\times10^{9}L_{\odot}$ and
their masses are typically a factor of 100 greater than
Galactic globular clusters (Scoville {\it et al.} 1999).
A likely explanation is that these objects are
associations of star clusters seen in more nearby
systems (Surace {\it et al.} 1998), since the typical
intercluster separation in nearby starburst regions is
too small ($\leq 18\,\mathrm{pc}$, see Meurer 1995) to be
resolved at high redshift. Since both such star
cluster associations and galactic nuclei may appear as
local brightness enhancements on ULIRG images and the
luminosities of the largest associations can be comparable
with those of galactic nuclei, it is necessary to find a
feasible criterion to distinguish these two types of
objects of different physical nature. Hereafter for
convenience, we adopt a nomenclature in which the luminous
star-forming regions alone are named as {\it knots} or
{\it star cluster associations}, while they are named as
{\it clumps} together with galactic nuclei.

Although there exists observational evidence for multiple
merging from investigations on both individual targets
and large samples, most of these works are based on
qualitative analysis and use somewhat subjective
criteria. Therefore, it is important to investigate
this issue on a quantitative basis; this becomes more
requisite considering the possibility of mistaking star
cluster associations as galactic nuclei in ULIRGs, since
they are similar in many aspects. This comprises the
main subject of the present paper; in our work, we
perform a careful photometric analysis on a large {\it HST}
snapshot imaging survey sample of ULIRGs (Borne {\it et al.}
1999a, 1999b), in order to find evidence for multiple merger
based primarily on a quantitative criterion of {\it I}-band
luminosity. On the other hand, there are still no
detailed comparisons of multiple mergers with other
populations of ULIRGs (e.g., double mergers and
single nucleus galaxies, see Sec. 5).
This provides the second motivation for our paper.
The structure of this paper is as follows: Sec. 2
includes basic information about the ULIRG sample
and their observations. In Sec. 3, we describe
details of our data reduction. In Sec. 4, we detail
our quantitative criterion for identifying putative
galactic nuclei. In Sec. 5, we present our main results, i.e.,
evidence for multiple mergers and a comparison of different
ULIRG populations. Finally, we discuss and summarize our
results in Sec. 6. Throughout this paper we assume
an Einstein-de~Sitter cosmology ($\Omega_0=1$) and adopt
$H_0=75\,\rm{km\,s^{-1}\,Mpc^{-1}}$. At the typical
redshift ($z=0.1$) of the sample galaxies, $0.1\arcsec$
(about 1 pixel) corresponds to roughly 200\,pc.

\section{Sample and observations}

For our purpose, we use archive data from an {\it HST}
snapshot imaging survey of ULIRGs (Borne {\it et al.}
1999a, 1999b). The sample images were taken using the
{\it Wide Field Planetary Camera 2} (WFPC2) in the
{\it I}-band pass (F814W), including 120 targets all
centered in the WF3 chip, of which the resolution is
0.0996\arcsec per pixel and the FOV is $800\times 800$
pixels. For each target in this survey, two 400{\rm s}
exposures were taken. Most of the observed targets
were selected from the QDOT sample (Leech {\it et al.}
1994; Lawrence {\it et al.} 1999), the rest were
selected from the brightest samples of Sanders {\it
et al.} (1998a, 1998b) and Melnick \& Mirabel (1990).
Apart from some sources which are saturated, mis-focused
or contaminated by bright foreground stars, we make
use of 97 independent targets for our study. In addition,
all these sources are selected to possess projected
nuclear separations less than 20\,kpc (for galaxies with
more than two putative nuclei, only minimum nuclear separations
are considered). This is because generally speaking, {\it
strong interactions} between galaxies are induced in
systems with nuclear separations below this value, i.e., 
the distance adequate for two interacting galaxies to
obtain a substantial disturbance down to their nuclear
radii of 1-2\,kpc within the lifetime of the encounter
(e.g., Lin {\it et al.} 1988). Basic information for
all the targets is listed in Table 1 in the order of
increasing {\it RA}, including {\it HST} visit No.,
target name, {\it RA}, {\it Dec}, redshift, far-infrared
luminosity and our classification (see Sec. 5). The
redshifts of these objects range from 0.04 to 0.35.

\placetable{tb1}

\section{Data reduction}

The sample images were first preprocessed through the
{\it standard Space Telescope Science Institute} (STScI)
pipeline, using a standard WFPC2-specific calibration
algorithm and the best available calibration files
(Holtzman {\it et al.} 1995). The preprocessing consists
of mask and analog-to-digital correction, bias and dark
subtraction, as well as flat-fielding, etc.. Our
post-pipeline calibrations include detection and removal
of warm pixels and cosmic ray events. The additional
steps of data reduction are discussed below.

\subsection{Selection for clumpy structures}

We use the standard {\it IRAF} task {\it imexamine} to
carefully browse the sources in the ULIRG sample. For
each target, three plots are displayed and compared
with each other -- the contour plot, the surface plot
and the snapshot image. The image is adjusted to
different greyscale levels until it renders the best 
visualization. The surface plot is always checked
from several different angles to get a full view. 
A set of ceiling values are tried until the contour plots
yield a clear view of the weak structures of the targets. By
careful comparison, common local brightness enhancements
appearing on each plot are marked as putative clumps.
On surface plots, such clump candidates are identified
on the basis of detectable brightness peaks; while on
contour plots, they are identified on the basis of
closed isophotal contours. As for snapshot images, it
is hard to resolve them by greyscale adjustment to
present all the clumps detectable in surface or contour
plots, except for those very bright or compact clumps
with high signal-to-noise ratio. However, snapshot
images are especially useful in that interacting
structures in ULIRGs are more perceptible in these
plots, and such structures are very useful for
determining whether or not several distinct separate
regions near the target position belong to a single
dynamically related system. This is done based on
the existence of detectable tidal streamers connecting
separate regions. Otherwise, we identify the object
nearest to the IRAS position as the target we need,
and its nearby objects are considered to be either
foreground or background ones (see also Sec. 6.1).
Hence our method will give a (conservative) lower limit
on the fraction of multiple merger objects.

In order to distinguish true physical clumps from
Poisson fluctuation peaks in the images, we make
rough FWHM (full width at half maximum) measurements
on the marginally detectable brightness peaks, and 
fluctuations are identified to be those peaks with
FWHMs smaller than the typical value of observed PSF
(point spread function) of {\it HST} WF3 chip with
F814W filter. Furthermore, it can not be ruled out
that foreground stars may coincidently fall onto
the target and are thus mistaken as ULIRG clumps.
A rough estimation of this probability is made using
all the 7 ULIRGs in the sample at relatively low
galactic latitude (i.e., $|b|<15^{\circ}$).
We use the standard {\it IRAF} task {\it daofind}
to search for star-like objects in these 7 ULIRGs which
are brighter than the minimum clump flux we have
identified in the sample (about 22.5\,mag). And by
multiplying the star number by the ratio of the
target area to the whole field area, we obtain an
average contamination number of about 0.1 for one
galaxy, hence the influence of star contamination
on our results is negligible. Here the area occupied
by each target is determined by counting the number
of pixels with ADU values at least $3\sigma$ above
the sky background. This is a conservative upper
limit, since the sub-sample we treat is contaminated
by foreground stars to the greatest extent due to
low galactic latitude. 

\subsection{Surface photometry}

We perform surface photometric measurements on the selected 
clumps to determine their {\it I}-band fluxes and 
luminosities. For well-separated clumps, we apply
standard aperture photometry programs in which
aperture widths are set to different values according
to their actual sizes, ranging from 0.2\arcsec\,
to 1.0\arcsec. The size is estimated from the
surface brightness profile when it reaches an
approximately constant value, i.e., the local background value.
In addition, estimates are made of the underlying
background galaxy flux by using the mean of the
pixels in a 1-pixel annulus immediately outside the
photometric aperture. However, some clumps in ULIRGs
are so close that their profiles overlap notably,
hence standard procedures do not function well.
In such cases we perform surface photometry in the
following way: we apply a large aperture size to
encircle all the overlapping clumps and the magnitude,
$m$, measured this way is considered to be the
combined {\it I}-band magnitude of encircled objects;
at the same time we apply a small aperture size
to measure the magnitude, $m'$, of the protrudent
part (i.e., non-overlapping region) of each
encircled clump, and the flux ratio of these
parts is approximated as that of the corresponding
true clumps, assuming that these overlapping clumps
possess similar profiles. Using $m$, $m_{1}'$ and
$m_{2}'$, we can estimate the true magnitude of
clumps from the following expression,
\begin{equation}
m_{1}=m+2.5\,\log\left(1+10^{\frac{m_{1}'-m_{2}'}{2.5}}\right),
\end{equation}
\begin{equation}
m_{2}=m+2.5\,\log\left(1+10^{\frac{m_{2}'-m_{1}'}{2.5}}\right).
\end{equation}
\noindent  
In all our surface photometry, the photometric 
calibration is performed using published
photometric solutions for WFPC2 and F814W
filter (Holtzman {\it et al.} 1995).

\placetable{tb2}

Simulations are performed in order to estimate the
uncertainty of the measured fluxes for overlapping
clumps. Using the standard {\it IRAF} task {\it
mkobjects}, we automatically produce two objects
(artificial clumps) with different flux ratios and
separations. Their FWHMs are fixed to be 5\,pixels.
For convenience, only de~Vaucouleurs brightness
profiles are considered. In addition, Poisson noise
is added to the simulated objects and the effects
of detector gain, readout and PSF are all modeled
using the appropriate parameters for the {\it HST}
WF3 chip. Table 2 lists details of each simulation
run. From the table, we can see that the largest
photometric uncertainty occurs when the separations
between clumps are small and the corresponding
magnitude difference of simulated clumps is high.
Since there are no overlapping clumps in this sample
with very different {\it I}-band magnitude, the
typical photometric uncertainty introduced by our
technique is less than 0.2\,mag, and the largest
uncertainty in our measurements, corresponding to
run 12 in Table 2, is around 0.5\,mag. Hence the
introduced photometric uncertainty does not affect
our results significantly.

\section{Identification of putative nuclei}

As mentioned in Sec. 1, one of the remarkable 
features of ULIRGs is that there always exist many
kpc or sub-kpc scale luminous clumpy structures;
and the key issue in our work is to distinguish
putative nuclei from star cluster associations.
It is found that typical masses for both galactic
nuclei in ULIRGs and the giant elliptical cores
are several times of $10^{9}M_{\odot}$ (Sakomoto
{\it et al.} 1999; Lauer 1985). This is typically
larger than the mass found for star formation
knots. In detailed studies of nearby starburst
galaxies such as M82 (de Grijs {\it et al.} 1999),
no star-forming knots have been observed to possess
masses as high as $10^{9}M_{\odot}$. And for the
nearest ULIRG, Arp~220, its most massive knot was
found to possess a mass of $6.6\times10^{8}M_{\odot}$
(Shioya {\it et al.} 2000). On the theoretical
side, complexes with masses greater than
$10^{9}M_{\odot}$ may suffer from disintegration
(Noguchi 1999). In addition, Taniguchi {\it et al.}
(1998) suggested that a large mass of $10^{9}M_{\odot}$
corresponds to a maximum star formation efficiency
of 1 for knots evolving from superclouds in
gravitationally unstable gas disks. Since the real
star formation efficiency must be smaller (e.g.,
0.1 from observations of the Galaxy), observed knots
are always less massive than $10^{9}M_{\odot}$.
Given these, we conservatively adopt
$1\times10^{9}M_{\odot}$ as the lower mass limit
to pick out putative nuclei from bright clumps in
the first step.

To express the above threshold in an observational
quantity, the lower mass limit is converted to
{\it B}-band luminosity using a constant mass-to-light
ratio of 6.5 appropriate for spheroids (Fugukita
{\it et al.} 1998). And the resultant value of
$1.54\times10^{8}L_{\odot}$ corresponds to
$M_{\rm{B}}=-15.0$. Adopting the statistical results
of Surace {\it et al.} (1998) that the typical $B-I$
color index for putative nuclei in ULIRGs is 2.0 with
a root mean square of 0.7, we obtain the lower
$M_{\rm{I}}$ limit of $-17.0$\,mag as a quantitative
criterion to pick out putative nuclei from bright clumps.
This is consistent with the argument that bright knots
may have an upper luminosity limit for intense
star-formation in a star cluster (Hutchings 1995).

We caution that this quantitative criterion may not be
sufficient to pick out ULIRG nuclei in some cases.
This is mainly due to two reasons. First,
the brightest knots and relatively faint nuclei
in ULIRGs may have comparable luminosities.
Second, the nuclear and circumnuclear regions of
ULIRGs are heavily obscured -- as a result,
a single nucleus may emerge as multiple
condensations due to complex obscuration patterns.
In order to eliminate the influences of these two
effects, we take further morphological considerations,
which are discussed below.

\subsection{Morphological considerations}

Generally speaking, in order to distinguish those
extremely bright knots from the putative nuclei,
color information is needed (e.g., Surace {\it et
al.} 1998, Scoville {\it et al.} 1999). However, we
do not incorporate color index in this paper,
because the ULIRGs imaging survey sample we used was
performed only in the {\it I}-band. In addition, the
results of $B-I$ color show that this index covers
a rather wide range from $-$0.3 to 5.2 for star-forming
knots (Surace {\it et al.} 1998). Considering the
relatively narrow span of 0.5 to 3.1 for putative
nuclei (Surace {\it et al.} 1998), it is obvious that
knots can be either bluer or redder than nuclei, as
was also pointed out by Hutchings (1995). Therefore,
the color index is not an ideal indicator to
distinguish putative nuclei from star formation knots.

On the other hand, although the luminosities of
some super star clusters are comparable to or even
greater than those of putative galactic nuclei, this
luminosity overlap is relatively insignificant compared
with the overlap of color index. Among all the 57 knots
in Surace's sample, less than 30\% are brighter than
our {\it I}-band luminosity threshold of $-$17.0\,mag
(see Fig 1 for the magnitude distribution of all these
57 knots). Furthermore, we suspect that some of these 
very bright knots in Surace's sample are possible
galactic nuclei (see Sec. 6.1).

Star-forming knots with large {\it I}-band
luminosities (i.e., comparable to those of putative nuclei)
may also exist in the current ULIRG sample. To rule 
them out on a single wavelength basis, we take into account the
dynamical information provided by interacting signatures of the
targets. This technique led Lipari {\it et al.} (1999) to detect
three galactic nuclei in NGC~3256 based on {\it HST} and
{\it ESO-NTT} observations. Therefore, after applying the
quantitative criterion of $M_{\rm{I}}$, we carefully examine
the local (such as nuclear arms and disks) as well as global
(such as tidal tails, plumes, rings) environments of the sample
targets to confirm whether or not clumps brighter than
$M_{\rm{I}}=-17.0$ are actual descendents of
progenitor bulges. Both numerical simulations and
observations show that each piece of such signatures
is connected to at least one nucleus residing at its
starting position (e.g., Toomre \& Toomre 1972;
Wright 1972). Because of this, some very bright
clumps (i.e., brighter than $-17.0$\,mag) are still
treated as massive star-forming regions if they are
located in the outer parts of the sample targets (e.g.,
at the tips of tail structures). Such objects are
more likely to be tidal galaxies produced by merging
(Fellhauer \& Kroupa 2000), hence they are not included
in our work. In our sample, there are 22 such clumps
brighter than $-17.0$\,mag in the {\it I}-band which
are removed due to these morphological considerations. Compared
with the total number of putative nuclei of nearly
200, this step, although important, should not affect our
statistical results.

This morphological consideration works fairly
well, since in most of our cases, such interacting
signatures are easy to identify and their
geometrical connections to bright clumps in
the same images are explicit. Thus
this technique  eliminates at least some of the uncertainties 
in our identification of putative nuclei due to the luminosity
overlaps between nuclei and knots. However,
it should still be notified that some levels
of subjectivity cannot be avoided in the
morphological considerations mainly due to two
reasons. The first is that faint interacting
signatures may be buried in the background noise
and are not detectable, especially for the targets
at relatively high redshift. The second is that the
starting positions of these structures are unclear
in some cases. e.g., it is not clear how to pick out the putative 
nuclei from a group of clumps located along a
tidal ring, purely from morphological signatures.
Since some features of interacting systems may be
due to quirks of pre-encounter disks and do not
lead to fundamental insights (Barnes 1999), it
is not surprising that we sometimes have
ambiguous circumstances. In such cases, we simply
apply our quantitative magnitude criterion
to identify putative nuclei.

\subsection{Obscuration effects}

Since the high far-infrared luminosity of ULIRGs results
from thermal re-radiation from dust (Sanders \&
Mirabel 1996), there is no doubt that ULIRG morphologies
tend to be affected by dust obscuration.
Images taken at longer wavelength are needed to examine
the extent to which the morphologies of the sample
targets are influenced. For the near infrared images
of 9 LIGs plus 15 ULIRGs taken by HST NICMOS
camera (at 1.1, 1.6 and 2.2 $\mu$m), it was stated that
although new interacting features as well as new super
star clusters could be found at longer wavelengths,
no new nuclei were seen in these images (Scoville {\it
et al.} 1999). This fact suggests that global morphologies
are not greatly affected by
dust obscuration effects  at optical wavelengths, and hence
our identification of putative nuclei should be reasonably secure.

In fact, to further minimize the above-mentioned effect,
some targets in the sample were treated separately, as in
the cases of \ufivezero, \usevenfour and \ubzero, etc.
Although these targets show signs of separate nucleus
components, the combined shapes of their brightness profiles
and loss of any nearby signatures of on-going interactions
suggest that they are more likely to be a single nucleus
split by a foreground dust lane. Surace {\it et al.} (1998)
have reported such cases in their warm ULIRG sample, e.g.,
IR05189$-$2524. This procedure is also consistent with
Borne {\it et al.}'s criterion that only cases with clearly 
separated optically luminous galactic components were
selected (Borne {\it et al.} 2000).

To summarize, the identification of putative galactic nuclei in our 
reduction includes two steps. The first step is  to apply directly an 
{\it I}-band luminosity criterion, in which all the ULIRG clumps 
brighter than $-$17.0\,mag were considered as nucleus candidates
for further consideration. The second step is to
include morphological information (primarily, the existence and
characteristics of tidal features), in order to remove some super
star clusters with {\it I}-band luminosities comparable
to those of putative nuclei. Only nuclei which pass through these two
criteria are considered as putative galactic nuclei. In the next section,
we present statistical analyses of this group of objects.

\section{Results}

Based on the quantitative criterion discussed in
Sec. 4 as well as morphological structures, we
pick out all the putative galactic nuclei from
clumpy structures in each sample galaxy. And all
the 97 ULIRGs can be classified as three categories
according to the number of their nuclei:

\begin{itemize}
\item Multiple nuclei ULIRGs (multiple mergers)
\item Double nuclei ULIRGs (double/pair mergers)
\item Single nucleus ULIRGs (single remnants)
\end{itemize}

\noindent
Note that for the double nuclei category, two
galaxies with projected nuclear separations of more
than 20\,kpc are excluded (IR13156$+$0435 and
IR15168$+$0045). Hence these three categories
compose a nearly complete sample of {\it strong
interacting systems and their remnants with
ultraluminous far-infrared luminosities}. Note
that here we adopt a working assumption that 
every nucleus present in merging systems
corresponds to a single progenitor nucleus 
(bulge), i.e., we assume that dynamical processes 
(such as tidal disruption) cannot produce compact 
regions with masses beyond $10^9M_\odot$. We 
caution that this assumption, while reasonable, 
needs to be verified using numerical simulations. 

Photometric results of all the putative nuclei
in the sample are listed in Tables 3, 4 and 5
for the three categories, respectively, including
positions and {\it I}-band absolute magnitudes.
In all these tables, nucleus positions are expressed
by {\it RA} and {\it Dec} relative to the corresponding
target positions given in Table 1, which
are fixed to pixel coordinate of (420,\,424) in
the {\it HST} WF3 chip. For multiple and double
mergers (see Tables 3 and 4), projected nuclear
separations as well as luminosity ratios are
included, and we only give minimum
separations for the multiple nuclei category.
Furthermore, some basic information about the
targets are also listed in these tables, including
tail lengths, brief comments on morphology and
available spectral types.

Our results show that among all the 97 ULIRGs of the
treated sample, 17 are multiple nuclei systems, 38 are
double nuclei systems, while the remaining 42 have only
one identifiable galactic nucleus. This gives plausible
fractions of ULIRGs with multiple, double and single
nucleus/nuclei as 18\%, 39\% and 43\%, respectively.
These results evidently support the multiple merger
scenario as a possible origin of ULIRGs, in addition
to the widely-accepted pair merger picture. This
argument was put forward by Borne {\it et al.} (2000)
based on the same sample. However, our results are
obtained in a more objective way due to our
mainly quantitative criterion. Comparisons
between Borne's results and ours are presented in
Sec. 6.2. In addition, further statistical studies
on this large sample reveal some 
differences among the three ULIRG categories in our
taxonomy, some of which give hints to a possible
evolutionary sequence from multiple mergers to double
mergers and then to single merger remnants (hereafter
M$\rightarrow$D$\rightarrow$S sequence, see Sec.
5.2). These results not only support our classification
scheme, but also give some insights into the dynamics
of galaxy interactions/mergers (see Sec. 6.2).

Besides putative nuclei, there often exist several
bright knots in most of the sample galaxies. One of
the significant characteristics of these bright knots is
that they tend to distribute around nuclear regions,
or in regions between/among separate nuclei. This
is consistent with the argument that star-forming
concentrations are preferentially situated along
the overlapping areas of interacting galactic disks
(Scoville {\it et al.} 1999). Besides these nuclear
knots, there are also cases with circum-nuclear knots,
or even with knots residing at the tips of tidal tails.
As pointed out by Weilbacher {\it et al.} (2000), a
possible descendent of such objects at the tips of
tidal tails is tidal dwarf
galaxies. Another tendency is that bright knots are
less frequently seen in single nucleus ULIRGs,
compared with multiple/double mergers. We plan to
study the properties of these bright knots
and investigate their formation mechanisms and
evolutionary fates in a future work.

\placetable{tb3}
\placetable{tb4}
\placetable{tb5}

\subsection{Multiple nuclei ULIRGs}

Table 3 gives the information on each nucleus
for all the 17 multiple mergers, while the {\it HST}
{\it I}-band images, surface plots and contour plots
are shown in Figs. 1-17 (arranged from left
to right, respectively). We put plus marks on each
image to indicate the positions of putative nuclei,
and the orientation of the image is indicated by the
headed north arrow and the unheaded east arrow. Since
pixel spacings are not always the same for the two
axes, the north and east directions may sometimes be
oblique. The orientation of each contour plot is
selected to be the same as the snapshot image for
direct comparison. For the three-dimensional surface
plot, we rotate it along both horizontal and vertical
viewing directions until it renders the best sight to
view each nucleus clearly. In addition, for each
multiple merger candidate, the snapshot image and
the contour plot both contain distance scale rulers,
and the scale of the surface plot is selected to
be the same as that of the contour plot. For the 17
multiple merger systems, the morphologies, the relative
positions and numbers of putative nuclei as well as
knots are very different. We give a description of each
target in turn.

\begin{description}
\item[\bf \udzero]
This galaxy is mainly composed of two parts separated
by 4.3\,kpc. The northeastern part contains a single
nucleus, while the southwestern part consists of two
close nuclei with a projected separation of only
570\,pc. Bright knots of the system preferentially
distribute to the southwest of each nucleus. There is a
45\,kpc tidal tail starting from the northeastern nucleus
and curving to the south. In addition, the contour plot
of \udzero presents distinct arm structures
connected to both parts of the system. These structures
may be remnants of progenitor spiral arms which have
not been completely destroyed by the merging process.

\item[\bf \uzerotwo]
The western part of \uzerotwo contains the brightest
nucleus of the system, while the eastern part is
composed of two close nuclei separated by 540\,pc.
The projected distance between the two parts is
about 2.6\,kpc and several bright knots can 
clearly be seen between them. A 50\,kpc tidal tail
starts at the eastern nucleus pair and curves from
the east to the west. A faint plume structure can
also be seen starting from the western nucleus and
stretching to the southwest. A possible evolutionary
scenario is that two spirals merge first and
produce the long tail we see; before
they finally evolve into a single object, a third
galaxy approaches from their western side, which
makes up a system with three putative nuclei.

\item[\bf \usixeight]
The morphology of this galaxy is like that of Mrk~273.
A very long tidal tail of nearly 40\,kpc starts from
the central star-like object and stretches from the northeast
to the southwest. Surface and contour plots of the
target clearly show that the central object is in fact
composed of three closely separated nuclei all brighter
than $-20.5$\,mag in the {\it I}-band. According
to the spectral information, the central active galactic
nuclei (AGN) phenomenon is
triggered in this target. The three progenitor galaxies
of this system are likely to have begun merging almost
simultaneously.

\item[\bf \usixnine]
The morphology of this galaxy is asymmetric and complex.
Three bright putative nuclei of about $-19.0$\,mag exist in the
system, with a roughly linear distribution. The projected
separation is about 800\,pc (between the center and
western nucleus) and 1.1\,kpc (between the center and
eastern nucleus). Two plume structures are
detectable in the system, with the western one being 9\,kpc
long and the northeastern one 5\,kpc long. Several
bright knots distribute just along these two plumes.
Furthermore, two much fainter plumes can also be seen from
the snapshot image, stretching to the southwest and east,
respectively. These fainter plumes are also much longer,
the southwestern one is nearly 28\,kpc long while the
eastern one is about 23\,kpc long.

\item[\bf \utwoseven]
 From the snapshot image, this galaxy
consists of two parts located to the northwest and
southeast, respectively. The projected separation
between these two parts is about 6.5\,kpc. The
southeastern component embeds two close nuclei
separated by 2.2\,kpc, and a tidal plume starts here
stretching as far as 10\,kpc. This gives further
support to the merger origin of the southeastern
component.

\item[\bf \utwoeight]
This galaxy is a system with peculiar morphology.
The three putative nuclei we identified approximately
distribute along a line. The central nucleus is
0.5\,kpc and 1\,kpc away from the eastern and
western nucleus, respectively. In the global
environment, the three closely separated nuclei
reside at the southern edge of a tidal ring with
11\,kpc diameter, and several bright knots are scattered
along the ring perimeter. These putative nuclei are also
connected with a 14\,kpc tail curving from the
east to the west. A possible explanation of this
peculiar morphology is that two galaxies interact
to produce a tidal tail, and before their bulges
finally merge together, a third galaxy impact on
them perpendicular to their disk planes, which
results in the ring structure we see now.

\item[\bf \uthreezero]
This system contains two main parts separated by
about 15\,kpc, and a material stream can be detected
connecting these two parts. The center of the
western part is a compact nucleus with {\it I}-band
absolute magnitude about $-19.3\,\rm{mag}$. Two
distinct spiral arms extend from it and stretch to
the north and south, respectively. Along the two
arms there are several bright and compact knots,
two of which are almost as bright as $-18.0$\,mag.
However, since these very bright clumps reside at
the tip of the southern spiral arm,
they are more likely to be progenitors of tidal
dwarf galaxies, and hence are not identified as putative
nuclei in our work (see
Sec. 4). In addition, two other nuclei brighter
than $-19.0$\,mag are located along the two spiral arms.

\item[\bf \uthreetwo]
Surface and contour plots of \uthreetwo show
clearly that there are three distinct putative nuclei in the
system. A close pair separated by 1\,kpc is located
to the east, while a fainter nucleus is located to the
west. The projected separation between these two parts
is about 2.4\,kpc. Although its three nuclei are closely
distributed, \uthreetwo only presents moderate
interacting features. A 10\,kpc arm starts from the
two eastern nuclei and curves to the southwest, while a
6\,kpc arm starts from the western nucleus and curves
to the north. Several bright knots preferentially distribute
along these two arms as well as around the two eastern
nuclei.

\item[\bf \ufivefour]
This is an object with very peculiar morphology which
may be hard to explain as a pair merger product. 
The snapshot image shows a round center
component, which is connected with two other bar-like
structures to the northwest and southwest, respectively.
Each of these three components encompasses a bright
nucleus. In fact, there are two closely separated clumps
in the southwestern {\it bar} which are both brighter
than $-18.0$\,mag. However, due to their contour shapes,
these two clumps appear to be a single nucleus
split by a foreground dust lane (as in the case of
IR05189$-$2524, see Surace {\it et al.} 1998). Hence
we conservatively take them as a single object, and the
$M_{\rm{I}}$ value listed in Table 3 is the
combined magnitude of these two objects.
A 11\,kpc tidal plume starts from the central component
and curves to the south. In addition, another tidal
tail as long as 19\,kpc is also detectable connected
to the northwestern {\it bar} and stretches from the
west to the east.

\item[\bf \ufourfour]
The morphology of this galaxy is rather complicated.
A strong tidal ring can be seen from the
{\it HST} snapshot image and its diameter is about
20\,kpc. The main part of this galaxy is located
at the northern edge of the ring perimeter. The galaxy
is clearly composed
of four compact nuclei, as shown in the surface and
contour plots. These four putative nuclei are very
bright, each with {\it I}-band absolute magnitude around
$-20.0\,\mathrm{mag}$. 6\,kpc away to the north of
these four nuclei is located the fifth nucleus of the
system, and a 12\,kpc plume structure just starts here
stretching from the south to the north.

\item[\bf \uninezero]
This is an example of a QSO hosted in an interacting
galaxy system. Besides the central brightest nucleus
with {\it I}-band absolute magnitude of nearly
$-23.0$, there exist another two bright nuclei which
are separated by 1.1\,kpc and located to the southeast.
A distinct ring structure with a diameter
of about 19\,kpc locates to the north of the system. 
This galaxy resembles 3C~48 and
IR04505$-$2958 in morphology and it provides us with
direct evidence that the AGN phenomenon can be triggered
by galaxy merging.

\item[\bf \uoneseven]
 From the HST morphology, this is clearly a
merging pair with two parts to the north and south,
respectively. These two parts are separated by
6\,kpc and some bright knots can be seen located
around each part. Both the northern and southern
parts consist of two distinct nuclei, separated by
1.6\,kpc and 3.1\,kpc, respectively. In addition, tidal
plume structures can be seen connected to these two
parts. From its morphological features, it is
likely that the progenitor of this system is a group
of four galaxies; each pair of galaxies merge first,
while the whole system begins its encounter before each
pair finishes its own merging process.

\item[\bf \uninethree]
 From the {\it HST} snapshot image, this 
galaxy is composed of two main parts to the south
and north, respectively. The projected separation
between these two parts is about 5\,kpc. The 
southern part encompasses a single nucleus, while
the northern part has four close nuclei in a
roughly linear distribution, whose high {\it I}-band
luminosities suggest that they are not very likely
to be star-forming knots. Only weak interacting
features are detectable in the system. A faint plume
stretching from the south to the north is connected
with the four northern nuclei. In addition, another
faint tidal ring is detectable to the east of
the main body.

\item[\bf \utwozero]
This ULIRG has a rather peculiar morphology which
consists of two parts from the snapshot
image. The northeastern part has a ring-like
structure with a diameter of more than 8\,kpc. Four
separate nuclei are concentrated at its south side, while
a number of bright knots can also be clearly seen
along the ring. The southwestern part of this system
consists of two close nuclei separated by
600\,pc and their connecting line is perpendicular
to the northeastern ring plane. In addition, there
exists a tidal bridge connecting these two parts.

\item[\bf \ufthree]
This is a three nuclei ULIRG with very peculiar
morphology. The bright object located in the center
is composed of two close nuclei separated by 1.2\,kpc,
while another much fainter nucleus resides to their 
northeast. A clear tidal bridge can be seen connecting
these two parts, and several bright knots are distributed
just along the bridge. Two plume structures of length
6\,kpc and 18\,kpc start from the central nucleus
pair and curve in opposite directions. In addition,
connected to the northeastern nucleus, there is
a 25\,kpc tidal tail stretching directly from the
east to the west.

\item[\bf \utwoone]
This system contains three putative nuclei with
very different luminosities. The one located at
the center has {\it I}-band luminosity as high as
$2.6\times10^{9}L_{\odot}$. The nucleus located
1.2\,kpc to the southeast of the central bright
nucleus and the one 2.4\,kpc to the west
are much fainter. In addition, bright star-forming
knots of the system are preferentially distributed
around the western nucleus. A multiple merger
scenario of this system is also supported by its
complex morphology with a 11\,kpc tidal tail and
two other plume structures.

\item[\bf \ufnine]
 From the snapshot image, this is a
galaxy with a relatively simple and regular morphology.
A distinct tidal tail about 19\,kpc long starts from
the central bright object and stretches from the northwest
to the southeast. The central object is in fact composed
of three close nuclei all brighter than $-19.5$\,mag,
and around them there are several bright star-forming
knots of the system. The three progenitor galaxies of
this system are likely to have begun to merge almost simultaneously.

\end{description}

\noindent
As described above, all of the targets listed in Table
3 encompass at least three putative nuclei and most of them
clearly show signs of strong interaction activities. Such
large scale tidal features coined as {\it tidal traumas}
(Barnes \& Hernquist 1992) include tails, bridges, rings,
plumes, etc. In some systems with very peculiar
morphologies, different types of tidal features may co-exist,
such as several tails or plumes (e.g., \ufthree in
Fig. 15). Borne {\it et al.} (2000) suggested that
merging pairs of spirals alone can not account for all the
dynamical diversities of ULIRGs, and numerical simulations
also support that such complex structures tend to appear
in multiple merger systems (Barnes \& Hernquist 1996).

\subsection{{\it I}-band magnitudes of putative nuclei and spectral types}

The {\it I}-band magnitude distributions of putative
nuclei for the whole sample and the three different
categories are presented in Figs. 18a-d, respectively.
The faint side cutoff in Figs. 18b and 18c (multiple
and double nuclei galaxies) is set by our criterion.
In Fig. 18d (single nucleus galaxies),
the faint side cutoff is at $-18.0$\,mag. This might be
due to the merger-induced formation of massive and bright
bulges in these systems. On the other hand, there exists
a distinct high luminosity tail above $M_{\mathrm{I}}=-21.0$
in Fig. 18d. This corresponds to a Seyfert~1/QSO
excess in single nucleus ULIRGs, since most nuclei
brighter than $-21$\,mag (9 out of 13 with
available spectra) belong to galaxies with Seyfert~1/QSO
spectra (see Tables 3, 4 and 5). In addition, 
Seyfert~1/QSO nuclei can occasionally be found in
multiple/double mergers (e.g., \uninezero and
\ufiveseven), and there is either likelihood that these
bright nuclei are primordial objects before merger or in
fact evolutionary descendents of previous merging steps.

The magnitude distributions of multiple/double mergers
and single nucleus galaxies give a median {\it I}-band
magnitude of $-19.3$ ($2.3\times 10^{9}L_{\odot}$) and
$-20.1$ ($4.7\times 10^{9}L_{\odot}$), respectively.
The high median luminosity for the single nucleus
category is partly due to the merger-induced formation
of massive, bright bulges, and partly due to the
triggering of active galactic nuclei (AGN) phenomenon
in a substantial fraction of merger remnants, as
mentioned above. In addition, these three categories
do not show much difference in the $M_{\mathrm{I}}$
distribution width. Although magnitude distributions
of multiple and double mergers do not present much
differences, it is obvious from the comparison of Figs.
18b and 18c that bright nuclei (i.e., from $-19.5$
to $-20.5\,\mathrm{mag}$) tend to appear more often
in double nuclei systems (nearly twice as often
than in multiple mergers). This gives some hints to
a possible evolutionary sequence from galaxy groups
to merger remnants with a single nucleus, along which
the double nuclei systems remain as intermediate
products (M$\rightarrow$D$\rightarrow$S sequence).

To check this M$\rightarrow$D$\rightarrow$S merger
sequence, we calculate the luminosity ratio of each
nucleus to the brightest one in the same target. Our
results show that nearly 40\% of the double merger
ULIRGs contain two nuclei with mass ratio more than
3:1 (assuming a constant mass-to-light ratio for
bulge). Although spiral galaxies might possess rather
different bulge-to-disk mass ratios, it cannot be
ruled out that a substantial fraction of putative nuclei in
pair merger are likely to be merger remnants of previous
galaxy encounters, since ULIRGs are postulated to arise
from mergers of two comparably massive gas-rich spiral
galaxies (e.g., Mihos 1999). This gives further
support to a possible M$\rightarrow$D$\rightarrow$S
evolutionary sequence.

Further investigation based on the optical spectra
available in the literature (Lawrence {\it et al.}
1999; Veilleux {\it et al.} 1999) as well as the
information from NED \footnote {The NASA/IPAC {\it
Extragalactic Database} (NED) is operated by the
Jet Propulsion Laboratory, California Institute
of Technology, under contract with the National
Aeronautics and Space Administration.} and our
observations shows that about 64\% (7/11) Seyfert~1/QSO
in this {\it HST} snapshot imaging survey sample belong
to galaxies with a single nucleus. This gives an important
hint that the Seyfert~1/QSO phenomenon tend to appear
at the final stage of galaxy interactions/mergers
when separate nuclei have merged together. Table
6 gives the detailed statistics of different
spectral types for the three categories. It shows
that the proportions of \ion{H}{2} region-like spectra
are 64\% (9/14), 35\% (8/21) and 20\% (5/25) for
multiple, double and single nucleus/nuclei galaxies,
respectively. While the AGN proportions for these
three categories are 36\% (5/14), 65\% (13/21) and
80\% (20/25). This clear trend of gradually changing
proportions of different spectral types suggests that
the M$\rightarrow$D$\rightarrow$S sequence is also
one with changing energetics: from starburst-dominated to
central AGN-dominated. Of course,
it should be mentioned that the \ion{H}{2}-like spectra
in some objects may be due to the super star clusters
rather than the identified putative nuclei, which is consistent
with the fact that the images of these targets are
distributed by knotty structures in their circumnuclear
regions (e.g., \uthreezero and \ufourfour).

\placetable{tb6}

\subsection{Nuclear separations and tail lengths}

Tables 3 and 4 list the projected nuclear
separations for multiple and double nuclei ULIRGs.
We present the corresponding distributions in
Fig. 19a. Given the {\it HST} WFPC2 resolution
limit, our measurements cannot probe separations
below several hundred parsecs, except for some low
redshift galaxies. At the maximum redshift in the
sample, $z=0.35$, our measurements can only probe
nuclear separations larger than 1.5\,kpc.
Therefore the distribution of nuclear separations
less than this value is incomplete. This incompleteness
is particularly serious at separations below 0.5\,kpc, since
very few targets in this sample are at low redshift
(i.e., less than 20\% with redshift smaller than
0.1). This is shown by the sharp decrease below
500\,pc in Fig. 19a. We also give the same
distribution based on results from the literature
(Surace {\it et al} 1999; Evans 1999; Rigopoulou
{\it et al.} 1999; Murphy {\it et al.} 1996;
Clements {\it et al.} 1996) in Fig. 19(b) for
comparison. One significant feature of our results
is that there is a high fraction (3/4) of nucleus
pairs with projected separations less than 5\,kpc,
which is much higher than the result from the
literature (1/3). The reason for this large difference
is mostly due to the low resolution limit of
ground-based observations in previous works. This
result supports that systems at late merging stage
preferentially host ULIRG galaxies (Mihos 1999).

Fig. 20 displays the correlation between nuclear
separations and tail lengths. Most of the data
points are distributed in the lower left conner of the
plot. However, considering that both the minimum
separations and tail lengths applied here are
projected values, the distribution profile of
data points is more instructive. Fig. 20 shows
a weak trend of smaller nuclear separations
coupled to longer tidal tails. Simple qualitative
analysis suggests a picture that when galaxies
first encounter at a relatively large distance,
their mutual tidal field is not strong enough to
produce long tail structures; and when they are
drawn closer as the merging process proceeds, disk
stars in each galaxy are catapulted farther out
and tidal tails continue to lengthen. This
picture is consistent with numerical simulations
(e.g., Toomre \& Toomre 1972; Wright 1972).
However, there do exist several examples with
both long tails and large nuclear separations
(e.g., \uninesix). This might imply that at
least one nucleus in such systems is in fact an
evolutionary descendent of a previous merging
process, or merely that the resolution limit of {\it HST}
prevents us from probing fine structures at its
very center. Considering this, the multiple merger
fractions we determined at the beginning of Sec. 5
might be an under-estimate of the true fraction. In addition,
relatively short tail structures can occasionally
be found in single nucleus ULIRGs. This further
strengthens the case that these systems are merger remnants
of previous galaxy encounters. Of course, our
measurements of tail length carry an obvious
bias: the observable tidal tails at a given 
surface brightness in distant galaxies tend to
be shorter than their nearby counterparts even 
if they have the same physical lengths,
due to the $(1+z)^4$ surface brightness dimming.
However, since the sample targets do not cover a
wide range of redshifts (0.04$\rightarrow$0.35),
this effect should not influence our statistical
results significantly.

\section{Discussion and summary}

\subsection{Multiple merger fraction}

Based on our quantitative criterion as well as morphological
features, we obtain the fractions for multiple, double and
single nucleus/nuclei ULIRGs as 18\%, 39\% and 43\%,
respectively. A question is why there is no report of
multiple merger cases in other sample studies on ULIRGs,
e.g., in the sample of Surace {\it et al.} (1998). To test
the consistency between these two results, we directly apply 
an upper mass limit of $10^{9}M_{\odot}$ on
their results of ULIRG knots as well as putative nuclei,
and find that two targets in their sample (IR12071$-$0444
and IR15206+3342) should be classified as multiple mergers,
namely, there are at least three clumps too massive to be
explained as star cluster associations in each of these
two targets. Considering this, the fraction of multiple 
nuclei ULIRGs in Surace's sample is 22\%, very close to 
our result of 18\%. The main reason why Surace
{\it et al.} (1998) did not choose these objects as
multiple merger systems might be that only the one or two
brightest clumps in their sample were believed to plausibly
represent putative nuclei. This means the pair merger
origin was considered to be the only
evolutionary picture for ULIRGs. This assumption may
sometimes give rise to some inconsistencies or ambiguities.
This is especially true in the case of IR12071$-$0444. For this system, 
there are three bright objects that have comparable masses (several
times $10^{9}M_{\odot}$), but only one of them was
selected as a putative nucleus. Furthermore,
this identified nucleus also possesses
$M_{\mathrm{B}}$, $M_{\mathrm{I}}$ and $B-I$ values
just between those of the other two bright clumps
(see Table 2 of Surace {\it et al.} 1998). This
means that at least this putative nucleus does
not show distinguished features compared with some other 
clumps in the same target. Thus it is possible that
any of the three brightest clumps in IR12071$-$0444
represent putative nuclei. On the other hand, in our work, 
the putative nuclei were identified independent of the evolutionary
origin for ULIRGs, in order to check the possibility of
the multiple merger scenario proposed by some researchers.

Based on the same sample, the multiple merger
candidates selected by Borne
{\it et al.} (2000) include 9 ULIRGs classified as
multiple nuclei galaxies and 8 ULIRGs classified
as interacting groups (three multiple mergers in
Borne's classification are absent from our subsample, and are
therefore not discussed here). The {\it HST} 
snapshot imaging survey sample does include a fraction of
ULIRGs which are very likely to be located within galaxy 
groups from their images, since there
are several bright or dwarf galaxies surrounding them.
However, we did not consider them as multiple merger
systems except when the central target contains at 
least three putative nuclei or is connected with some
of its companions by obvious interacting signatures.
This is because further redshift observations need to
be performed for confirmation. These group candidates
include \udtwo, \udfive, \useventhree, \utwofive\, and
\uffive. In fact, we have carried out some spectroscopic
observations using the 2.16\,m telescope at the Xinglong
Station of Beijing Astronomical Observatory and identified
several luminous/ultraluminous IRAS galaxies located
in galaxy groups (Zou {\it et al.} 1995;
Wu {\it et al.} 1998). However, for two multi-merger
(group) candidates in Table 1
(\ufzero\, and \ueighttwo), neither of them belong to 
an interacting group since the surrounding galaxies
are either foreground or background objects (Zheng 
{\it et al.} 1999). Considering that the evolutionary
sequence from galaxy groups to ULIRGs and then to
ellipticals is attractive from both theoretical
and observational points of view, further detailed 
multi-wavelength observations on these group 
candidates have been scheduled.

As for the remaining 12 multiple mergers in Borne's
classifications, six of them are consistent with our
classifications (\utwoseven, \ufivefour, \ufourfour,
\uoneseven, \uninethree and \ufnine). We carefully
examined the six discrepant targets, of which five
of them contain only one or two clumps brighter than
$-17.0\,\mathrm{mag}$ in the {\it I}-band. Therefore
they are classified as single or double nucleus/nuclei
ULIRGs. The case for \ueightfive\, is somewhat confusing.
This target seems to reside in a group composed of
several bright galaxies as well as star-like objects,
of which the biggest galaxy indeed contains three
obvious nuclei. However, the star-like object located
in the center is the nearest target to the IRAS
position. In addition, the QDOT redshift survey
(Lawrence {\it et al.} 1999) reveals that its optical
spectral type is Seyfert~1 and the calculation of
far-infrared luminosity for \ueightfive\, is also based
on the redshift of this Seyfert galaxy. Considering
this, Borne {\it et al.} (2000) may have identified
the wrong target on the WFPC2 snapshot image.
Putting aside \ueightfive, we suggest that the
remaining five multiple mergers in Borne's
identification which are absent from our results
are due to different selection criteria involved. 
Borne {\it et al.} (2000) identified several targets
with very complex tidal features to be multiple merger
cases with no regard for
the exact numbers of their observable nuclei. Although
it is likely that a very complex morphology does point
to a possible multiple merger origin, we identify
multiple mergers simply based on the exact number of
clumps brighter than $-17.0$\,mag, because any tight
correlation between morphologies and nucleus numbers
needs to be verified by further numerical simulations
of merging processes.

\subsection{Some hints for merging dynamics}

\begin{description}
\item[1.]
 From luminosity distributions in Sec. 5.2, it
is obvious that putative nuclei in single nucleus
ULIRGs tend to be more luminous than those in
multiple/double mergers. Statistical studies on
available spectral information also reveal that
there is a Seyfert~1/QSO excess in the single
nucleus category. Therefore, our results strongly
support the argument of Kauffmann \& Haehnelt (2000)
that bulge and supermassive black holes may both
grow in galaxy merging. And our results also give some
hints to a possible close relation between central black
holes and the bulges of their host galaxies.
(Gebhardt {\it et al.} 2000).

\item[2.]
Although only a small fraction of snapshot images are
given in this paper, there is no doubt from morphological
comments in Tables 3, 4 and 5 that most ULIRGs with
single nuclei have weak interacting signatures (in many
cases, only weak plume structures are detectable). On the
other hand, peculiar morphologies frequently emerge in
double/multiple nuclei ULIRGs, such as long tidal tails
and distinct ring structures. This indicates that strong
interacting features always thin out at final merging
phase, which is consistent with results from numerical
simulations (e.g., Barnes \& Hernquist 1996).

\item[3.]
 From the detailed descriptions in Sec. 5.1, we
can see that multiple merger processes may be very
complicated and the morphologies of their intermediate
products may be very diverse. This indicates that
the evolutionary history of interactions/mergers
in galaxy groups must vary due to different
initial conditions. When a group of galaxies begins
to merge, the central massive spiral may swallow
its satellite galaxies one by one; on the other hand,
merging processes may also happen between several galaxy
pairs or among sub-groups simultaneously, then these
separate parts start a second merging step. Major
mergers between comparably massive spirals alone
cannot account for the dynamical diversity of the
ULIRG population (Borne {\it et al.} 2000).

\end{description}

\acknowledgments
We would like to express our gratitude to the staff members
at the Beijing Astronomical Center, especially W.-P. Lin
and H. Wu, for generous help in data reduction. We also wish
to thank Z. Zheng and the anonymous referee
for useful comments and suggestions, which
improved the paper. We are also grateful to
Martin Smith for a careful reading of the text.
This research was based on observations
obtained with the NASA/ESA {\it Hubble Space Telescope}
through the Space Telescope Science Institute, Astronomy,
Inc., under NASA contract NAS5-26555. This project was
supported by the NSFC and NKBRSF G19990754.

\clearpage
\begin{deluxetable}{ccrrccl}
\footnotesize
\tablecaption{Sample of ULIRGs observed with WFPC2 of HST. \label{tb1}}
\tablewidth{0pt}
\tablehead{
\colhead{No.}&\colhead{Target name}&\colhead{RA(2000)}&\colhead{Dec(2000)}&
\colhead{Redshift$^{a}$}&\colhead{$\log \frac{L_{\mathrm{fir}}}{L_{\odot}}^{b}$}&
\colhead{Classification}
}
\startdata
uc7 &\ucseven      &00:08:38   &$-$15:26:52   &0.195   &   12.0   &Double\nl
uc9 &\ucnine       &00:13:04   &$-$01:23:05   &0.164   &   11.9   &Double\nl
u33 &\uthreethree  &00:17:45   &   49:54:11   &0.149   &   11.9   &Double\nl
ud0 &\udzero       &00:18:43   &$-$08:33:36   &0.109   &   11.8   &Multiple\nl
u57 &\ufiveseven   &00:23:22   &   10:46:22   &0.230$^{*}$   &$<$12.3   &Double\nl
u59 &\ufivenine    &00:30:04   &$-$28:42:26   &0.280   &   12.4   &Single\nl
ud2 &\udtwo        &00:36:01   &$-$27:15:35   &0.069   &   11.8   &Single\nl
u60 &\usixzero     &00:48:39   &$-$07:12:19   &0.243   &   12.2   &Double\nl
u01 &\uzeroone     &00:53:35   &   12:41:36   &0.062   &   11.5   &Single\nl
ud4 &\udfour       &01:01:31   &$-$03:36:28   &0.176   &   12.0   &Double\nl
ud5 &\udfive       &01:05:37   &$-$22:39:18   &0.186$^{*}$   &   11.9   &Single\nl
u65 &\usixfive     &02:08:07   &   08:50:04   &0.345   &   12.5   &Single\nl
u02 &\uzerotwo     &02:38:13   &$-$47:38:12   &0.098   &   12.0   &Multiple\nl
uf0 &\ufzero       &02:48:28   &$-$02:21:35   &0.180   &   12.0   &Double\nl
u68 &\usixeight    &03:54:25   &$-$64:23:45   &0.300   &   12.6   &Multiple\nl
ub0 &\ubzero       &03:57:11   &$-$82:55:16   &0.140$^{*}$   &   11.8   &Single\nl
u69 &\usixnine     &04:39:51   &$-$48:43:15   &0.203   &   12.2   &Multiple\nl
u70 &\usevenzero   &04:44:31   &   26:14:10   &0.171   &   12.1   &Single\nl
u71 &\usevenone    &05:13:24   &$-$48:07:58   &0.162   &   12.0   &Double\nl
ub1 &\ubone        &05:19:12   &   77:48:12   &0.158   &   11.9   &Single\nl
u72 &\useventwo    &05:25:27   &$-$23:32:08   &0.171   &   11.9   &Single\nl
u03 &\uzerothree   &06:02:54   &$-$71:03:09   &0.079   &   12.0   &Double\nl
u04 &\uzerofour    &06:21:01   &$-$63:17:23   &0.091   &   12.1   &Double\nl
u73 &\useventhree  &06:30:13   &   35:07:50   &0.170   &   12.0   &Double\nl
u74 &\usevenfour   &06:36:36   &$-$62:20:32   &0.159   &   12.2   &Single\nl
ub4 &\ubfour       &06:51:46   &   22:04:30   &0.144   &   12.2   &Double\nl
u75 &\usevenfive   &06:59:06   &   18:58:21   &0.188   &   12.1   &Double\nl
ub5 &\ubfive       &07:29:12   &   61:18:53   &0.138   &   11.7   &Single\nl
u76 &\usevensix    &07:41:23   &   32:08:09   &0.170   &   11.9   &Single\nl
u77 &\usevenseven  &08:23:13   &   27:51:39   &0.167   &   12.1   &Double\nl
ub7 &\ubseven      &08:38:04   &   50:55:09   &0.097   &   11.8   &Double\nl
ub8 &\ubeight      &08:53:16   &$-$15:15:48   &0.135   &   12.0   &Single\nl
ub9 &\ubnine       &09:06:34   &   04:51:28   &0.124   &   11.9   &Double\nl
u05 &\uzerofive    &09:35:52   &   61:21:11   &0.040   &   11.9   &Single\nl
u49 &\ufournine    &09:45:21   &   17:37:54   &0.128   &   11.7   &Single\nl
u50 &\ufivezero    &09:45:29   &   19:15:50   &0.150$^{*}$   &   11.7   &Single\nl
u78 &\useveneight  &10:05:42   &   43:32:39   &0.177   &   11.9   &Single\nl
uc2 &\uctwo        &10:15:21   &   49:28:19   &0.154$^{*}$   &   11.9   &Single\nl
u79 &\usevennine   &10:58:39   &   38:29:06   &0.207   &   12.1   &Single\nl
u80 &\ueightzero   &11:00:34   &   04:22:08   &0.173   &   11.9   &Single\nl
uc3 &\ucthree      &11:11:37   &   53:34:57   &0.143   &   11.8   &Double\nl
u06 &\uzerosix     &11:12:03   &$-$02:54:23   &0.105   &   12.1   &Double\nl
u81 &\ueightone    &12:13:20   &   31:40:53   &0.207   &   12.2   &Single\nl
u07 &\uzeroseven   &12:13:46   &   02:48:41   &0.072   &   12.2   &Double\nl
u82 &\ueighttwo    &12:22:47   &   16:29:45   &0.181   &   12.1   &Single\nl
u34 &\uthreefour   &12:51:41   &$-$10:25:26   &0.100   &   11.8   &Double\nl
u24 &\utwofour     &13:16:54   &   23:40:46   &0.138   &   11.8   &Single\nl
u25 &\utwofive     &13:36:24   &   39:17:29   &0.180   &   12.2   &Single\nl
u83 &\ueightthree  &13:36:51   &   63:47:04   &0.237   &   12.4   &Double\nl
u08 &\uzeroeight   &13:44:42   &   55:53:11   &0.038   &   12.0   &Double\nl
u26 &\utwosix      &13:46:39   &   23:06:21   &0.142   &   12.1   &Single\nl
u36 &\uthreesix    &13:48:40   &   58:18:52   &0.158   &   12.1   &Double\nl
u27 &\utwoseven    &13:56:10   &   29:05:36   &0.109   &   11.9   &Multiple\nl
u28 &\utwoeight    &14:08:19   &   29:04:46   &0.117   &   11.9   &Multiple\nl
u29 &\utwonine     &14:18:59   &   45:32:12   &0.151   &   11.8   &Single\nl
u30 &\uthreezero   &14:22:31   &   26:02:06   &0.159   &   12.1   &Multiple\nl
u31 &\uthreeone    &14:33:28   &   28:11:59   &0.175   &   12.0   &Double\nl
u85 &\ueightfive   &14:36:58   &$-$41:47:11   &0.182   &   12.0   &Single\nl
u09 &\uzeronine    &14:37:38   &$-$15:00:23   &0.082   &   12.2   &Double\nl
u10 &\uonezero     &14:40:59   &$-$37:04:33   &0.068   &   12.0   &Single\nl
u32 &\uthreetwo    &14:59:37   &   32:44:58   &0.114   &   11.8   &Multiple\nl
u38 &\uthreeeight  &15:44:05   &$-$10:09:00   &0.160   &   11.8   &Single\nl
u54 &\ufivefour    &16:02:33   &   37:34:53   &0.185$^{*}$   &   11.6   &Multiple\nl
u86 &\ueightsix    &16:18:37   &$-$04:09:44   &0.213   &   12.2   &Single\nl
u39 &\uthreenine   &16:46:59   &   45:48:23   &0.191   &   12.2   &Single\nl
u41 &\ufourone     &16:55:20   &   52:56:36   &0.194   &   12.1   &Double\nl
u42 &\ufourtwo     &17:18:54   &   54:41:49   &0.148   &   12.1   &Single\nl
u87 &\ueightseven  &17:47:10   &$-$51:58:44   &0.175   &   12.0   &Single\nl
u88 &\ueighteight  &17:47:05   &   58:05:21   &0.310   &   12.4   &Single\nl
u44 &\ufourfour    &18:58:14   &   65:31:29   &0.177   &   12.0   &Multiple\nl
u14 &\uonefour     &19:31:22   &$-$72:39:20   &0.061   &   11.8   &Double\nl
u15 &\uonefive     &19:32:22   &$-$04:00:01   &0.086$^{*}$   &   12.2   &Single\nl
uc5 &\ucfive       &19:59:49   &$-$47:48:17   &0.139   &   11.9   &Double\nl
u90 &\uninezero    &20:06:31   &$-$15:39:06   &0.192   &   12.4   &Multiple\nl
u16 &\uonesix      &20:11:24   &$-$02:59:52   &0.106   &   12.3   &Multiple\nl
u17 &\uoneseven    &20:13:30   &$-$41:47:34   &0.130   &   12.5   &Multiple\nl
u91 &\unineone     &20:14:06   &$-$29:53:51   &0.143   &   11.8   &Single\nl
u92 &\uninetwo     &20:21:11   &$-$47:47:07   &0.178   &   12.1   &Single\nl
u93 &\uninethree   &20:28:38   &$-$37:47:09   &0.180   &   12.0   &Multiple\nl
u94 &\uninefour    &20:34:18   &$-$19:09:12   &0.153   &   11.9   &Double\nl
u18 &\uoneeight    &20:44:18   &$-$16:40:16   &0.088   &   12.1   &Single\nl
u95 &\uninefive    &20:54:26   &$-$54:01:17   &0.228   &   12.2   &Double\nl
u19 &\uonenine     &20:58:27   &$-$42:39:03   &0.043   &   11.9   &Double\nl
u20 &\utwozero     &21:16:19   &$-$44:33:40   &0.093   &   12.0   &Multiple\nl
u96 &\uninesix     &21:58:16   &$-$58:09:40   &0.165   &   12.0   &Double\nl
uf3 &\ufthree      &22:23:29   &$-$27:00:03   &0.132   &   12.1   &Multiple\nl
u21 &\utwoone      &22:51:49   &$-$17:52:24   &0.078   &   12.0   &Multiple\nl
uf5 &\uffive       &22:57:24   &$-$26:21:13   &0.164   &   11.9   &Double\nl
u22 &\utwotwo      &23:15:47   &$-$59:03:14   &0.044   &   11.8   &Double\nl
u97 &\unineseven   &23:16:35   &   04:05:17   &0.220   &   12.1   &Single\nl
uf6 &\ufsix        &23:17:14   &$-$11:00:37   &0.101   &   11.8   &Double\nl
u98 &\unineeight   &23:24:28   &   29:35:39   &0.241   &   12.3   &Double\nl
u23 &\utwothree    &23:26:04   &$-$69:10:19   &0.106   &   12.1   &Single\nl
uf7 &\ufseven      &23:26:50   &$-$03:41:06   &0.189$^{*}$   &   11.8   &Single\nl
u56 &\ufivesix     &23:39:01   &   36:21:08   &0.065   &   12.0   &Single\nl
uf8 &\ufeight      &23:43:40   &   02:45:04   &0.092   &   11.8   &Double\nl
uf9 &\ufnine       &23:54:10   &$-$24:04:25   &0.153   &   11.9   &Multiple\nl
\enddata
\tablenotetext{a}{Redshifts for most targets are from the PSCz catalogue, with
some exceptions from on-line NED database (marked by an asterisk).}
\tablenotetext{b}{Far infrared luminosities were calculated from the following
expression (Sanders \& Mirabel 1996) 
\begin{center}
$F_{\mathrm{fir}}=1.26\times 10^{-14}\{2.58f_{60}+f_{100}\}[\mathrm{W\,m^{-2}}],$\\
$L_{\mathrm{fir}}=L(40-500\mu\mathrm{m})=4\pi D_{\mathrm{L}}^{2}CF_{\mathrm{fir}}[L_{\odot}]$,
\end{center}
where $D_{\mathrm{L}}$ is the luminosity distance,
the scale factor $C$ is adopted as 1.6, and
$f_{60}$ and $f_{100}$ are the
60\,$\mathrm{\mu m}$ and 100\,$\mathrm{\mu m}$ fluxes
from the PSCz catalogue, respectively.
}
\end{deluxetable}

\clearpage
\begin{deluxetable}{ccccc}
\footnotesize
\tablecaption{Photometric uncertainties in simulated overlapping clumps.\label{tb2}}
\tablewidth{0pt}
\tablehead{
\colhead{Run No.$^a$}&\colhead{Actual $\Delta m_{\mathrm{I}}$ (mag)}&
\colhead{Separation (pixel)}&\colhead{Measured $\Delta m_{\mathrm{I}}$ (mag)}&
\colhead{uncertainty (mag)}
}
\startdata
1&0.0&8&0.14&0.08\nl
2&0.0&6&0.19&0.06\nl
3&0.0&4&0.05&0.03\nl
4&0.2&8&0.09&0.06\nl
5&0.2&6&0.38&0.08\nl
6&0.2&4&0.05&0.02\nl
7&0.5&8&0.72&0.17\nl
8&0.5&6&0.57&0.09\nl
9&0.5&4&0.30&0.07\nl
10&1.0&8&1.14&0.22\nl
11&1.0&6&0.85&0.01\nl
12&1.0&4&0.51&0.04\nl
\enddata
\tablenotetext{a}{Each parameter combination ($\Delta{m_I}$ and
separation) consists of ten independent realizations, in which
the magnitude of one artificial clump is selected to range from
16.0\,mag to 17.8\,mag, with a step of 0.2\,mag, and the magnitude
of the other artificial clump is determined by adding the
corresponding $\Delta m_{\mathrm{I}}$. Measured $\Delta m_{\mathrm{I}}$
is the mean value of these ten processes, and the uncertainty refers
to the corresponding root mean square error. In all the simulation
runs with different flux ratios, the two artificial
clumps are always too severely overlapped to be distinguished from each
other as their separation drops to around 3\,pixel, thus we choose
the minimum simulated separation to be 4\,pixel; on the other hand,
as the clump separation is larger than around 8\,pixel, the two
clumps in any of the realizations are so well-separated that their
fluxes can be determined precisely by applying standard aperture
photometry programs, thus the maximum simulated separation of
8\,pixel is chosen.}
\end{deluxetable}

\clearpage
\begin{deluxetable}{crrcccccc}
\scriptsize
\tablecaption{Properties of putative nuclei in multiple mergers.\label{tb3}}
\tablewidth{0pt}
\tablehead{
\colhead{Target name}&\colhead{$\Delta RA (\arcsec)$}&\colhead{$\Delta Dec (\arcsec)$}&
\colhead{$M_{\mathrm{I}}$ (mag)}&\colhead{$\frac{L}{L_{\mathrm{max}}}^a$}&
\colhead{$S_{\mathrm{min}}$ (kpc)$^b$}&\colhead{$L_{\mathrm{tail}}$ (kpc)$^c$}&
\colhead{Morphology}&\colhead{Spectral type}
}
\startdata
\udzero      &   0.12&   0.15&$-$17.95&0.40& 0.57&40&tail&LINER\nl
&$-$0.71&$-$0.48&$-$18.94&1.00&\nl
&$-$0.85&$-$0.46&$-$18.62&0.74\nl
\uzerotwo    &   0.22&$-$0.48&$-$19.10&1.00& 0.54&48&tail&\nl
&   0.86&$-$0.30&$-$18.61&0.64\nl
&   1.01&$-$0.28&$-$17.98&0.64\nl
\usixeight   &$-$0.27&   0.89&$-$22.37&1.00& 1.91&38&tail, plume&Sy2\nl
&$-$0.49&   0.80&$-$20.86&0.25\nl
&$-$0.77&   0.96&$-$21.16&0.33\nl
\usixnine    &   0.35&   0.47&$-$18.45&0.40& 0.85&&plume&\ion{H}{2}\nl
&   0.47&   0.45&$-$19.31&0.88\nl
&   0.65&   0.43&$-$19.45&1.00\nl
\utwoseven   &$-$0.16&$-$0.65&$-$17.95&0.08& 2.16&&plume&\ion{H}{2}\nl
&   1.21&$-$1.28&$-$18.96&0.20\nl
&   1.65&$-$0.95&$-$20.69&1.00\nl
\utwoeight   &$-$0.17&$-$1.17&$-$20.08&1.00& 0.51&12&tail, ring&\ion{H}{2}\nl
&   0.03&$-$1.06&$-$19.88&0.83\nl
&   0.15&$-$1.07&$-$18.80&0.31\nl
\uthreezero  &$-$0.17&$-$0.02&$-$18.83&0.65& 2.23&&arm&\ion{H}{2}\nl
&   0.16&   0.55&$-$19.30&1.00\nl
&   0.21&   0.20&$-$19.07&0.81\nl
&   2.70&$-$0.67&$-$19.05&0.79\nl
\uthreetwo   &   1.32&$-$1.14&$-$18.35&0.19& 1.01&&arm&\nl
&   1.77&$-$0.88&$-$20.16&1.00\nl
&   1.76&$-$0.63&$-$19.12&0.38\nl
\ufivefour   &   0.81&$-$1.26&$-$19.35&0.35& 8.13&18&tail, plume&\nl
&   0.44&$-$0.33&$-$20.49&1.00\nl
&   0.06&   0.69&$-$20.46&0.97\nl
\ufourfour   &$-$0.13&   1.49&$-$18.90&0.19& 0.95&&ring, plume&Sy2\nl
&   0.35&   0.78&$-$20.70&1.00\nl
&   0.21&   0.69&$-$19.83&0.45\nl
&   0.29&   0.52&$-$20.50&0.83\nl
&$-$0.05&   0.25&$-$19.69&0.39\nl
\uninezero   &$-$1.25&$-$0.51&$-$19.50&0.04& 1.12&&star-like, ring&QSO\nl
&$-$1.23&$-$0.70&$-$18.90&0.03\nl
&$-$0.25&$-$0.87&$-$22.90&1.00\nl
\uoneseven   &   1.87&$-$0.48&$-$17.48&0.28& 1.55&&plume&\ion{H}{2}\nl
&   1.29&$-$0.12&$-$18.12&0.50\nl
&   1.22&   1.24&$-$18.30&0.59\nl
&   0.90&   1.11&$-$18.87&1.00\nl
\uninethree  &   1.26&   0.56&$-$19.77&0.88& 0.64&&ring, plume&\ion{H}{2}\nl
&   0.88&   1.37&$-$18.57&0.29\nl
&   0.99&   1.47&$-$19.72&0.84\nl
&   1.09&   1.51&$-$19.91&1.00\nl
&   1.35&   1.67&$-$18.68&0.32\nl
\utwozero    &   1.10&$-$1.18&$-$18.46&0.14& 0.60&&ring&\ion{H}{2}\nl
&   0.80&$-$1.13&$-$17.96&0.09\nl
&   0.62&$-$1.27&$-$17.59&0.06\nl
&   0.44&$-$1.26&$-$18.26&0.12\nl
&$-$0.20&$-$2.05&$-$18.48&0.15\nl
&$-$0.14&$-$2.32&$-$20.56&1.00\nl
\ufthree     &   0.26&   0.35&$-$18.31&0.21& 1.17&25&tail, plume&\ion{H}{2}\nl
&$-$0.56&$-$1.24&$-$20.03&1.00\nl
&$-$0.81&$-$1.32&$-$19.83&0.83\nl
\utwoone     &$-$0.48&$-$1.00&$-$17.71&0.20& 1.22&11&tail, plume&\ion{H}{2}\nl
&$-$0.51&$-$0.72&$-$19.44&1.00\nl
&$-$1.32&$-$0.48&$-$17.95&0.25\nl
\ufnine      &$-$0.09&   0.15&$-$19.84&1.00& 0.92&16&tail&LINER\nl
&   0.30&   0.24&$-$19.53&0.75\nl
&   0.15&   0.35&$-$19.67&0.86\nl
\enddata
\tablenotetext{a}{The {\it I}-band luminosity ratio of each nucleus to
the brightest one in the same target.}
\tablenotetext{b}{The projected minimum nuclear separation.}
\tablenotetext{c}{The projected visible length of the tidal tail.}
\end{deluxetable}

\clearpage
\begin{deluxetable}{crrcccccc}
\scriptsize
\tablecaption{Properties of putative nuclei in pair mergers.\label{tb4}}
\tablewidth{0pt}
\tablehead{
\colhead{Target name}&\colhead{$\Delta RA (\arcsec)$}&\colhead{$\Delta Dec (\arcsec)$}&
\colhead{$M_{\mathrm{I}}$ (mag)}&\colhead{$\frac{L}{L_{\mathrm{max}}}^a$}&
\colhead{$S_{\mathrm{min}}$ (kpc)$^b$}&\colhead{$L_{\mathrm{tail}}$ (kpc)$^c$}&
\colhead{Morphology$^d$}&\colhead{Spectral type}
}
\startdata
\ucseven        &$-$0.48&   0.56&$-$20.04&0.85& 1.84&18&tail&\nl
&$-$0.32&   0.41&$-$20.22&1.00\nl
\ucnine         &$-$0.69&   0.21&$-$19.45&0.47& 4.47&17&tail, plume&\nl
&$-$1.51&   0.40&$-$20.27&1.00\nl
\uthreethree    &$-$0.62&   0.37&$-$19.47&1.00& 9.03&20&ring, tail&\ion{H}{2}\nl
&$-$0.53&$-$1.45&$-$17.16&0.12\nl
\ufiveseven     &   0.91&   1.23&$-$21.94&1.00& 1.92&11&tail&Sy1\nl
&   0.87&   0.95&$-$19.66&0.12\nl
\usixzero       &$-$1.00&   0.63&$-$20.37&1.00& 2.06&16&tail&\nl
&$-$1.03&   0.34&$-$18.41&0.16\nl
\udfour         &$-$1.38&   0.08&$-$20.33&1.00&13.83&17&tail&\nl
&   0.94&   0.90&$-$19.27&0.38\nl
\ufzero         &$-$0.92&$-$0.68&$-$20.49&1.00& 5.91&&plume&\nl
&$-$0.45&   0.24&$-$20.00&0.64\nl
\usevenone      &$-$0.50&$-$0.87&$-$18.47&1.00& 0.67&&plume&LINER\nl
&$-$0.55&$-$0.75&$-$17.80&0.54\nl
\uzerothree     &   1.24&   1.44&$-$19.69&0.74& 8.65&18&tail, plume&\nl
&$-$1.42&   0.13&$-$20.01&1.00\nl
\uzerofour      &   0.60&   0.16&$-$17.96&1.00& 3.76&18&tail&\nl
&   1.19&   1.13&$-$17.71&0.79\nl
\useventhree    &$-$0.58&   1.31&$-$19.40&0.54& 8.11&18&tail&\ion{H}{2}\nl
&   1.28&   0.00&$-$20.06&1.00\nl
\ubfour         &$-$1.44&   0.20&$-$19.59&0.81& 1.48&&plume&\ion{H}{2}\nl
&$-$1.72&   0.08&$-$19.82&1.00\nl
\usevenfive     &$-$0.63&   0.67&$-$20.49&1.00& 6.53&&plume&LINER\nl
&$-$0.41&$-$0.42&$-$19.34&0.35\nl
\usevenseven    &   0.78&$-$0.49&$-$20.06&1.00& 6.01&23&tail&\ion{H}{2}\nl
&   0.29&   0.51&$-$17.68&0.11\nl
\ubseven        &   0.22&   0.18&$-$19.03&1.00& 2.71&&ring, plume&\nl
&   0.84&   0.65&$-$17.69&0.29\nl
\ubnine         &$-$0.48&$-$0.28&$-$19.18&0.95& 1.07&28&tail, plume&LINER\nl
&$-$0.62&   0.28&$-$19.24&1.00\nl
\ucthree        &   1.06&$-$1.52&$-$19.62&0.69& 1.18&&ring, plume&Sy1\nl
&   1.25&$-$1.67&$-$20.03&1.00\nl
\uzerosix       &$-$0.42&   0.97&$-$18.54&1.00& 1.23&18&tail&LINER\nl
&$-$0.17&   1.19&$-$17.98&0.60\nl
\uzeroseven     &   0.08&$-$0.03&$-$18.66&1.00& 3.74&32&tail, bridge&\nl
&$-$0.85&$-$1.05&$-$17.45&0.33\nl
\uthreefour     &$-$1.05&$-$0.09&$-$19.87&1.00& 2.89&&ring, plume&\nl
&$-$0.30&   0.20&$-$18.08&0.19\nl
\ueightthree    &$-$0.75&$-$1.25&$-$18.64&0.64&11.49&19&tail&\nl
&   0.47&$-$0.13&$-$19.13&1.00\nl
\uzeroeight     &$-$0.07&$-$1.53&$-$18.15&1.00& 0.36&$>$31&tail&Sy2\nl
&   0.11&$-$1.37&$-$17.72&0.67\nl
\uthreesix      &   1.25&$-$0.33&$-$19.74&1.00& 3.97&&arm&\nl
&   0.50&$-$0.46&$-$17.57&0.14\nl
\uthreeone      &   1.60&$-$1.41&$-$20.78&1.00& 6.84&&ring, plume&Sy2\nl
&   0.93&$-$2.42&$-$18.55&0.13\nl
\uzeronine      &$-$1.03&$-$0.01&$-$19.31&1.00& 4.83&15&tail&LINER\nl
&$-$1.87&$-$1.36&$-$17.43&0.18\nl
\ufourone       &   1.00&   1.48&$-$20.13&1.00& 5.60&&&Sy2\nl
&   0.50&   0.70&$-$18.62&0.25\nl
\uonefour       &   0.23&   0.15&$-$20.17&1.00& 8.79&$>$28&tail&Sy2\nl
&   1.01&$-$3.55&$-$19.42&0.50\nl
\ucfive         &   0.39&   1.59&$-$18.94&0.45& 1.55&18&tail, plume&\nl
&   0.22&   1.32&$-$19.80&1.00\nl
\uonesix        &$-$0.17&$-$1.39&$-$17.49&0.81& 0.76&&plume&LINER\nl
&   0.04&$-$1.38&$-$17.72&1.00\nl
\uninefour      &$-$0.71&$-$1.56&$-$19.95&1.00& 4.98&&plume&\ion{H}{2}\nl
&$-$1.67&$-$1.77&$-$18.45&0.25\nl
\uninefive      &$-$0.67&   1.21&$-$20.25&1.00& 1.27&&&LINER\nl
&   0.20&$-$1.94&$-$20.23&0.98\nl
\uonenine       &$-$0.07&   2.14&$-$19.45&1.00& 0.88&25&tail, plume&\ion{H}{2}\nl
&$-$0.10&$-$2.66&$-$18.61&0.46\nl
\uninesix       &$-$1.04&$-$2.68&$-$19.40&0.77&14.01&30&tail, plume&LINER\nl
&   2.70&$-$2.44&$-$19.69&1.00\nl
\uffive         &   0.33&   0.84&$-$20.32&1.00& 4.67&&plume&\ion{H}{2}\nl
&$-$0.18&   0.13&$-$19.19&0.35\nl
\utwotwo        &   0.64&   0.10&$-$17.93&0.75& 3.79&17&tail, plume&\nl
&   0.80&$-$2.06&$-$18.24&1.00\nl
\ufsix          &   5.92&$-$1.77&$-$19.69&1.00& 1.77&54&tail&\nl
&   6.39&$-$1.86&$-$17.89&0.19\nl
\unineeight     &   0.39&   0.61&$-$18.64&0.13& 1.84&&&\ion{H}{2}\nl
&   0.20&   0.78&$-$20.85&1.00\nl
\ufeight        &$-$0.52&$-$0.87&$-$20.23&1.00& 2.78&&plume&Sy1\nl
&$-$1.33&$-$0.65&$-$17.51&0.08\nl
\enddata
\tablenotetext{a}{The {\it I}-band luminosity ratio of each nucleus to
the brightest one in the same target.}
\tablenotetext{b}{The projected minimum nuclear separation.}
\tablenotetext{c}{The projected visible length of the tidal tail
(tidal tails in \uzeroeight\, and \uonefour\, are beyond the FOV of the snapshot images,
therefore the tail lengths given here are lower limit values).}
\tablenotetext{d}{Blank line in this column refers to no detectable interacting signatures.}
\end{deluxetable}

\clearpage
\begin{deluxetable}{crrcccc}
\scriptsize
\tablecaption{Properties of putative nuclei in single nucleus galaxies.\label{tb5}}
\tablewidth{0pt}
\tablehead{
\colhead{Target name}&\colhead{$\Delta RA$ (\arcsec)}&\colhead{$\Delta Dec$ (\arcsec)}&
\colhead{$M_{\mathrm{I}}$ (mag)}&\colhead{$L_{\mathrm{tail}}$ (kpc)$^a$}&\colhead{Morphology$^b$}&
\colhead{Spectral type}
}
\startdata
\ufivenine      &$-$0.68&   0.00&$-$23.14& 8&tail&Sy1\nl
\udtwo          &   0.98&$-$1.37&$-$20.44&&arm&\nl
\uzeroone       &$-$1.00&$-$1.31&$-$21.67&&star-like, arm&Sy1\nl
\udfive         &   0.82&$-$0.47&$-$20.27&17&tail, plume&\nl
\usixfive       &$-$0.97&$-$1.22&$-$23.46&&star-like&Sy1\nl
\ubzero         &   0.31&$-$1.25&$-$21.75&&star-like, arm&Sy1\nl
\usevenzero     &   1.30&   1.11&$-$19.93&&arm&Sy2\nl
\ubone          &$-$0.28&$-$0.05&$-$19.77&&&\nl
\useventwo      &   0.40&   0.95&$-$20.02&&arm&\nl
\usevenfour     &   1.19&   0.53&$-$20.32&4&tail&\ion{H}{2}\nl
\ubfive         &$-$0.33&$-$1.17&$-$20.43&&&Sy2\nl
\usevensix      &$-$0.48&$-$0.06&$-$20.60&&&Sy2\nl
\ubeight        &$-$0.09&   1.03&$-$18.48&&arm&\nl
\uzerofive      &   1.08&$-$0.42&$-$18.31&&&Sy1.5\nl
\ufournine      &   1.08&   1.01&$-$20.78&&&Sy2\nl
\ufivezero      &   0.44&   0.91&$-$21.37&&star-like, plume&Sy1\nl
\useveneight    &   0.98&$-$0.31&$-$22.49&&star-like&Sy1\nl
\uctwo          &   0.89&   0.31&$-$18.23&&&LINER\nl
\usevennine     &   0.68&$-$0.20&$-$20.60&&plume&\ion{H}{2}\nl
\ueightzero     &$-$0.02&   0.40&$-$19.60&&plume&LINER\nl
\ueightone      &   1.35&$-$0.18&$-$19.98&&plume&\ion{H}{2}\nl
\ueighttwo      &   1.09&   0.19&$-$21.96&&&LINER\nl
\utwofour       &   0.41&$-$0.04&$-$19.51&&plume&\nl
\utwofive       &   0.08&$-$2.05&$-$22.60&&star-like, arm&Sy1\nl
\utwosix        &   1.75&   1.13&$-$18.65&&&\nl
\utwonine       &   0.84&$-$1.28&$-$20.01&&plume&\nl
\ueightfive     &$-$1.39&$-$0.69&$-$21.67&&star-like&\nl
\uonezero       &   3.51&$-$0.53&$-$19.41&&&Sy2\nl
\uthreeeight    &$-$0.50&$-$0.53&$-$19.75&&&\nl
\ueightsix      &   1.22&$-$0.43&$-$21.43&&&\nl
\uthreenine     &$-$0.72&$-$0.18&$-$19.94&&plume&\nl
\ufourtwo       &$-$0.97&$-$0.47&$-$20.03&&&\nl
\ueightseven    &   1.85&   1.63&$-$19.79&&&LINER\nl
\ueighteight    &$-$0.71&$-$0.93&$-$21.09&&plume&Sy2\nl
\uonefive       &$-$0.66&$-$0.30&$-$18.28&&plume&\ion{H}{2}\nl
\unineone       &   0.74&$-$0.70&$-$19.57&&plume&\nl
\uninetwo       &$-$1.37&   1.88&$-$19.42&&arm&\nl
\uoneeight      &$-$0.24&$-$0.66&$-$20.12&&&\ion{H}{2}\nl
\unineseven     &   0.05&$-$1.36&$-$21.66&&plume&LINER\nl
\utwothree      &$-$0.37&$-$1.18&$-$19.18&&plume&\nl
\ufseven        &$-$1.38&$-$0.88&$-$20.50&&arm&\nl
\ufivesix       &$-$0.73&$-$0.63&$-$18.15&&&LINER\nl
\enddata
\tablenotetext{a}{The projected visible length of the tidal tail.}
\tablenotetext{b}{Blank line in this column refers to regular elliptical morphology.}
\end{deluxetable}

\clearpage
\begin{deluxetable}{cccc}
\footnotesize
\tablecaption{Spectral information for different ULIRG populations.$^a$\label{tb6}}
\tablewidth{0pt}
\tablehead{
\colhead{Spectral type}&\colhead{Single remnants}&\colhead{Double mergers}&
\colhead{Multiple mergers}
}
\startdata
Seyfert~1/QSO&28\%&13\%&7\%\nl
Seyfert~2/LINER&52\%&52\%&29\%\nl
\ion{H}{2}&20\%&35\%&64\%\nl
\enddata
\tablenotetext{a}{The fraction of galaxies with available spectral
information is 82\% (14/17), 61\% (23/38) and 60\% (25/42) for multiple,
double and single nucleus/nuclei ULIRGs, respectively.}
\end{deluxetable}

\clearpage

\centerline{\Large\bf Figure captions}

\figcaption{{\it I}-band magnitude distribution of
57 star-forming knots in the warm ULIRG sample (Surace
{\it et al.} 1998), which are visually fainter than 
$22.5$\,mag (i.e., the minimum clump flux
in our reduction).
The bin width is 0.5\,mag. The data in the plot are 
cited from the literature.}

\figcaption{Snapshot image, surface and contour plots
for \udzero. Scale ruler is 5\,kpc in the snapshot image
and contour plot. This galaxy consists of two components,
of which the northeastern one contains one nucleus and
the southwestern one contains two closely separated
nuclei. A long tidal tail is clearly seen in the
snapshot image.}

\figcaption{Snapshot image, surface and contour plots
for \uzerotwo. Scale ruler is 15\,kpc in the snapshot
image and 5\,kpc in the contour plot. The nuclear region of this galaxy
mainly contains two parts, the western part encompasses
a bright nucleus and the eastern part encompasses a
close nucleus pair. A distinct tidal tail and a faint
plume structure can be seen in the system.}

\figcaption{Snapshot image, surface and contour plots
for \usixeight. Scale ruler is 15\,kpc in the snapshot image
and 5\,kpc in the contour plot. Morphological and spectral features
of this galaxy is like those of the well-known Mrk~273.
The central star-like object consists of three separate
nuclei, and is connected with a very long tidal tail.}

\figcaption{Snapshot image, surface and contour plots
for \usixnine. Scale ruler is 15\,kpc in the snapshot image
and 5\,kpc in the contour plot. This galaxy has a complex morphology
which consists of three bright nuclei in the center
and four plume structures with much different properties,
such as length, intensity, and alignment.}

\figcaption{Snapshot image, surface and contour plots
for \utwoseven. Scale ruler is 5\,kpc in the snapshot image
and contour plot. This is a galaxy with relatively
weak interacting features. The big, bright object
to the southeast is in fact composed of two separate
nuclei.}

\figcaption{Snapshot image, surface and contour plots
for \utwoeight. Scale ruler is 5\,kpc in the snapshot image
and 1\,kpc in the contour plot. Three putative nuclei closely distribute
in the central region of the system. Distinct tidal tail
as well as ring structure can be seen in the snapshot
image.}

\figcaption{Snapshot image, surface and contour plots
for \uthreezero. Scale ruler is 5\,kpc in the snapshot image
and contour plot. This system contains two parts, of
which the northwestern one consists of three separate
nuclei. Two arm structures can be seen clearly from
the snapshot image.
In contour and surface plots, we only display
the northwestern part of this system to get a clearer view.}

\figcaption{Snapshot image, surface and contour plots
for \uthreetwo. Scale ruler is 5\,kpc in the snapshot
image and 1\,kpc in the contour plot. This is a multiple
nuclei ULIRG with moderate interacting features.}

\figcaption{Snapshot image, surface and contour plots
for \ufivefour. Scale ruler is 10\,kpc in the snapshot
image and 5\,kpc in the contour plot. This galaxy
consists of three distinct
components, each of which encompasses a bright nucleus.
Tidal tail and plume structures can be clearly seen
in the system.}

\figcaption{Snapshot image, surface and contour plots
for \ufourfour. Scale ruler is 5\,kpc in the snapshot
image and 1\,kpc in the contour plot. This is a typical
example of interacting group with complex merging
history. The other three bright
objects along the tidal ring are foreground stars. The
faintest nucleus locates at the north. Although it is not
so compact as the other four putative nuclei, its {\it I}-band
absolute magnitude of nearly $-19.0$ suggests that it
is not very likely to be a star cluster association.}

\figcaption{Snapshot image, surface and contour plots
for \uninezero. Scale ruler is 5\,kpc in the snapshot image
and contour plot. This ULIRG is an example of QSO hosted
in an interacting galaxy system. Although the two
clumps to the northwest of the system are not very
compact, their high luminosities as well as the existence
of a distinct tidal ring suggest that they cannot be
explained as star-forming knots. Ceiling values are applied
in contour and surface plots to display clearly faint
structures of the system.}

\figcaption{Snapshot image, surface and contour plots
for \uoneseven. Scale ruler is 5\,kpc in the snapshot image
and contour plot. This galaxy contains two main parts,
each of which encompasses two close nuclei. Tidal
plumes can be seen connected with each part.}

\figcaption{Snapshot image, surface and contour plots
for \uninethree. Scale ruler is 5\,kpc in the snapshot image
and contour plot. The northern part contains a single
nucleus while the southern part is composed of four
separate nuclei. This is a system with relatively weak
interacting features.}

\figcaption{Snapshot image, surface and contour plots
for \utwozero. Scale ruler is 5\,kpc in the snapshot image
and contour plot. This is an interacting system with very
peculiar morphologies. Four putative nuclei distribute along
a ring structure, while the other two nuclei are away
from the ring plane.}

\figcaption{Snapshot image, surface and contour plots
for \ufthree. Scale ruler is 5\,kpc in the snapshot image
and contour plot. This galaxy mainly contains two parts
connected by a tidal bridge. The northeastern part
encompasses a single nucleus, while the southwestern
one encompasses a close nucleus pair. A straight tail
and two plume structures can be seen from the snapshot image.}

\figcaption{Snapshot image, surface and contour plots for
\utwoone. Scale ruler is 5\,kpc in the snapshot image and
1\,kpc in the contour plot. This is a merging system with
three putative nuclei,
and the central nucleus is much brighter than the other
two. In surface and contour plots, ceiling values
are applied to display clearly the structures of the
two faint nuclei, as well as star-forming knots.
Another bright object clearly seen from surface and contour
plots is more likely to be a foreground star.}

\figcaption{Snapshot image, surface and contour plots
for \ufnine. Scale ruler is 5\,kpc in the snapshot image
and contour plot. Three putative nuclei of this system
concentrate in the center and are very close to each
other. The tidal tail can be seen clearly stretching
to the southeast.}

\figcaption{{\it I}-band magnitude distribution
of putative nuclei for the whole sample and the
three categories, respectively, as indicated by
the labels. The bin width is $0.25\,\mathrm{mag}$
for each plot.}

\figcaption{Distribution of separations between putative
nuclei. The upper panel presents distributions of minimum
projected nuclear separations of our measurements, while
the lower panel presents results from the literature, in
which all the separation values have been re-scaled to 
$H_0=75\,\mathrm{km\,s^{-1}\,Mpc^{-1}}$. The bin width
is 500\,pc for each plot.}

\figcaption{Correlation between projected nuclear
separations and tidal lengths, in which targets
of different categories are marked with different
symbols. For multiple nuclei systems, minimum
nuclear separations are adopted. Several targets
with tidal tails extending beyond the image FOV are
excluded here.}

\clearpage
\setcounter{figure}{0}

\begin{figure}
\centerline{\epsfig{figure=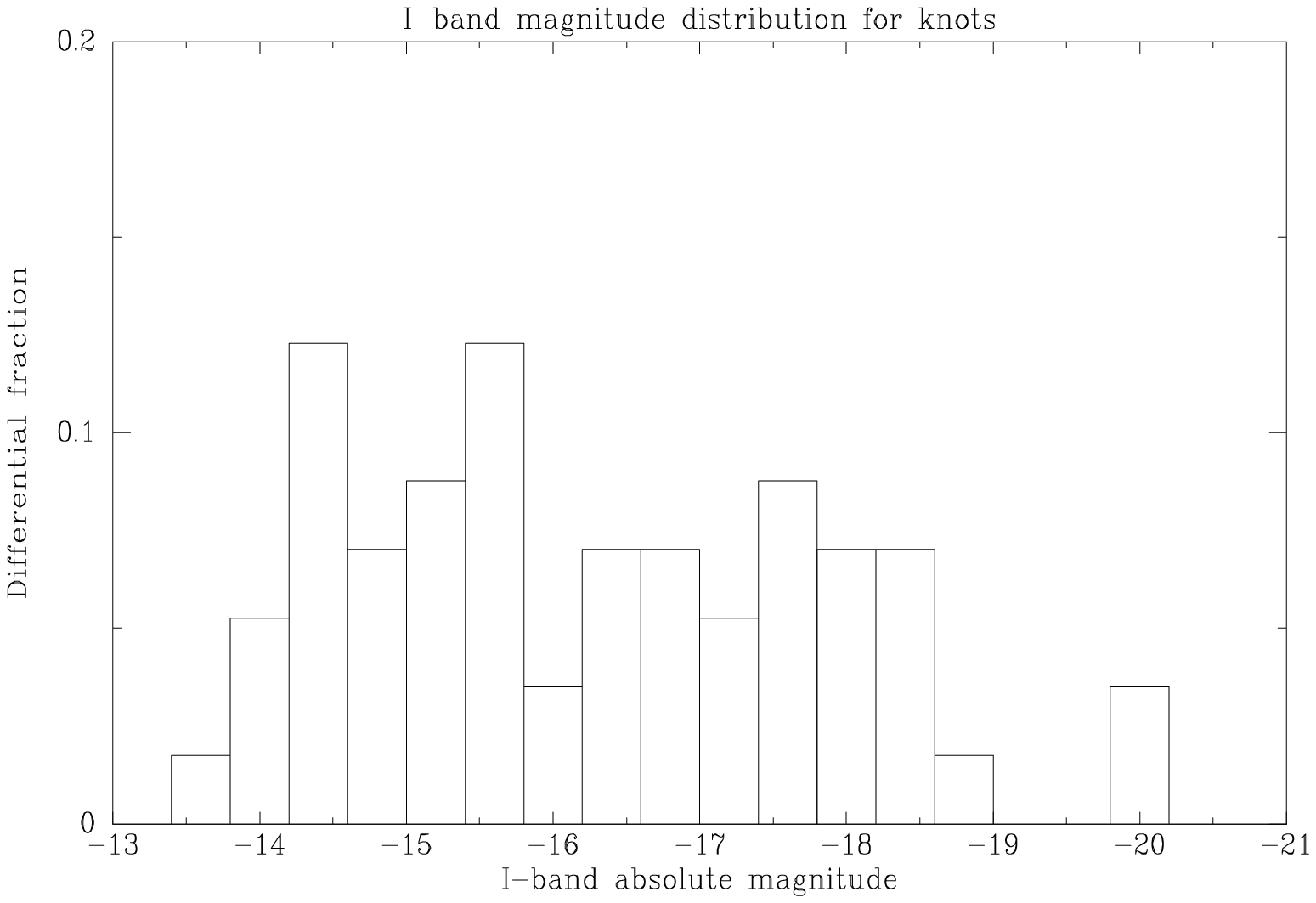,width=7cm}}
\caption[]{}\label{fig1}
\end{figure}

\begin{figure}
\centerline{\epsfig{figure=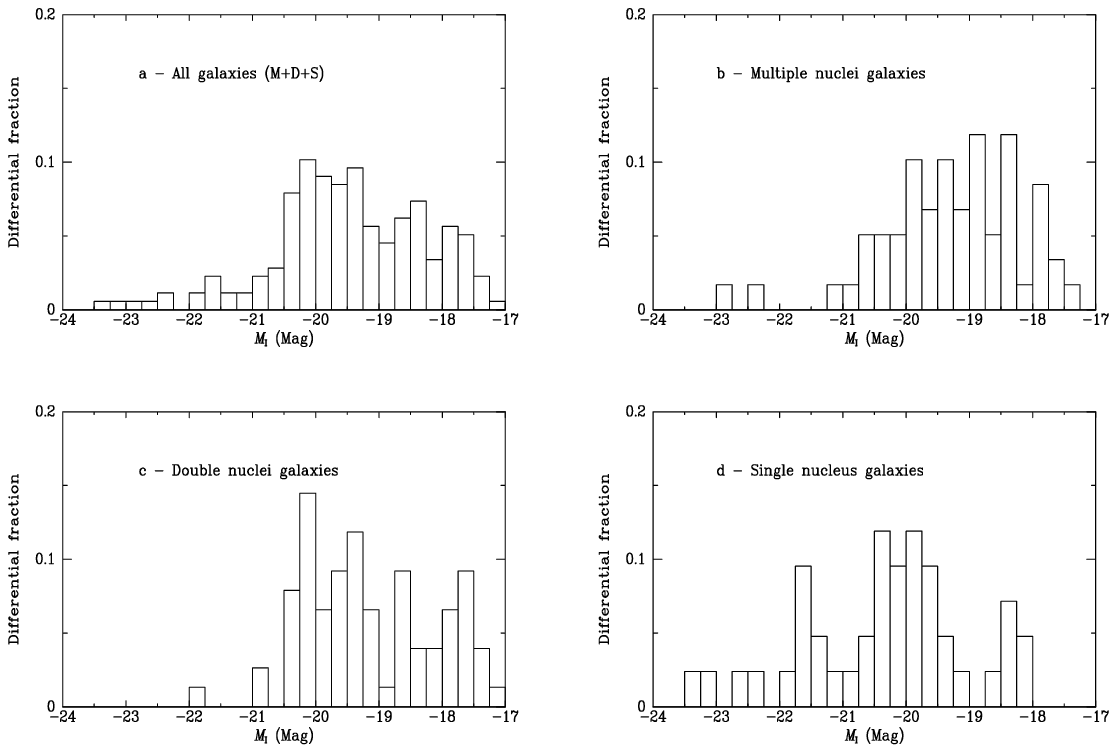,width=14cm}}
\caption[]{}\label{fig19}
\end{figure}

\begin{figure}
\centerline{\epsfig{figure=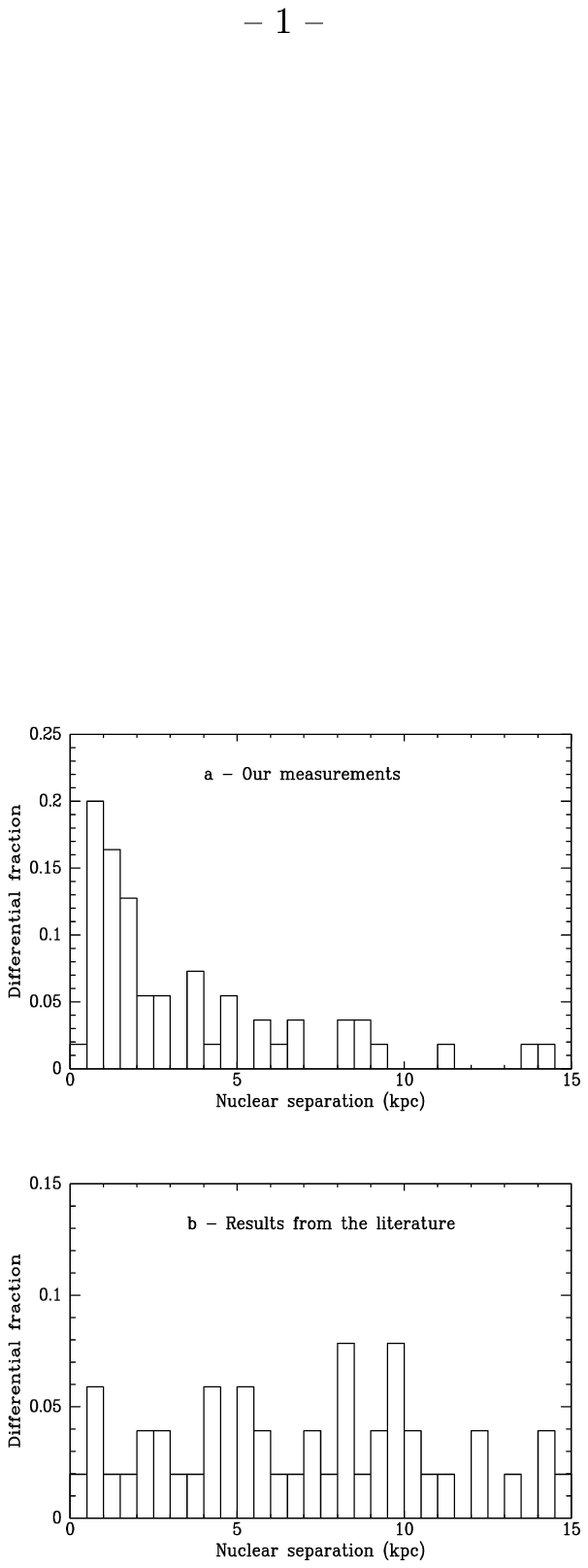,width=7cm}}
\caption[]{}\label{fig20}
\end{figure}

\begin{figure}
\centerline{\epsfig{figure=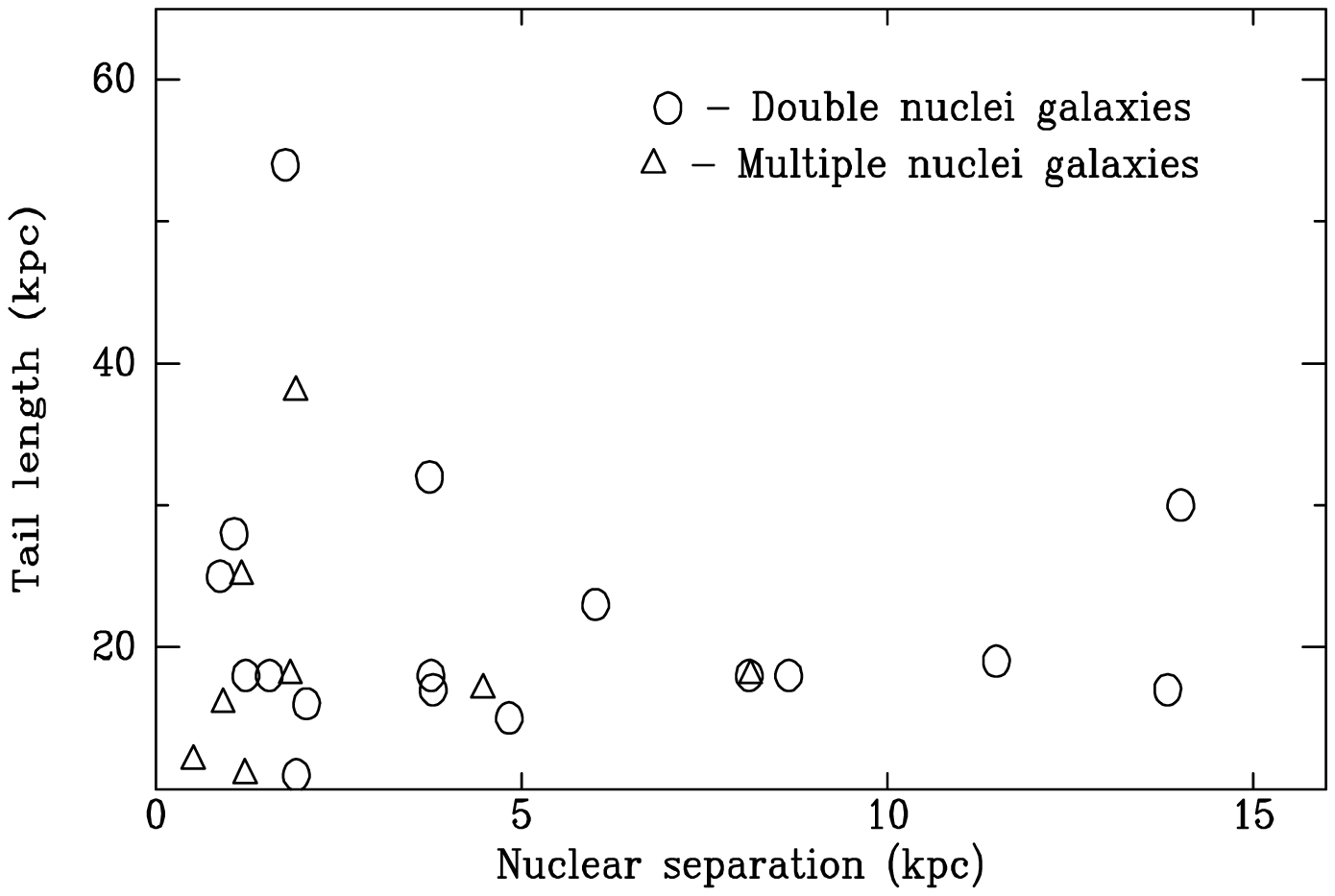,width=7cm}}
\caption[]{}\label{fig21}
\end{figure}

\end{document}